\documentclass[fleqn,usenatbib]{mnras}

\usepackage{newtxtext,newtxmath}
\usepackage[T1]{fontenc}
\DeclareRobustCommand{\VAN}[3]{#2}
\let\VANthebibliography\thebibliography
\def\thebibliography{\DeclareRobustCommand{\VAN}[3]{##3}\VANthebibliography}
\usepackage{graphicx}	% Including figure files
\usepackage{amsmath}	% Advanced maths commands
\usepackage{latexsym}
\usepackage{physics}
\usepackage{subfigure}

\title[MHD winds]{Accretion of supersonic magnetized winds onto black holes}

\author[Gracia-Linares and \& Guzm\'an]{
Miguel Gracia-Linares\thanks{E-mail: mgracia@austin.utexas.edu}${}^{1}$
and Francisco S. Guzm\'an\thanks{E-mail: francisco.s.guzman@umich.mx}${}^{2}$
\\
${}^{1}$ Center for Gravitational Physics,
Department of Physics. 
The University of Texas at Austin
Austin, TX 78712, U. S. A.\\
${}^{2}$Instituto de F\'{\i}sica y Matem\'{a}ticas, Universidad
              Michoacana de San Nicol\'as de Hidalgo. Edificio C-3, Cd.
              Universitaria, 58040 Morelia, Michoac\'{a}n,
              M\'{e}xico.
}

\date{Accepted XXX. Received YYY; in original form ZZZ}

\pubyear{20XX}

\begin{document}
\label{firstpage}
\pagerange{\pageref{firstpage}--\pageref{lastpage}}
\maketitle

% Abstract of the paper
\begin{abstract}
We present the accretion of magnetized supersonic winds onto a rotating black hole in three dimensions. We select representative spin-wind orientations in order to illustrate its effects on the evolution and morphology of the shock cone. The most important finding in the magnetized case, unlike the purely hydrodynamical scenario, is the formation of rarified spots where the magnetic field pressure dominates over the gas pressure. In these rarified spots we find the formation of eddies within the shock cone.
\end{abstract}


%     ---------------------------------------------
%     ----->     INTRODUCTION     <-----
%     ---------------------------------------------

\section{Introduction}
\label{sec:Introduction}

The Bondi-Hoyle-Littleton (BHL) or wind accretion occurs when a compact object moves through a constant fluid which is considered to be perfect and free of self-gravity, or conversely that a uniform fluid moves toward the accretor \citep{Bondi1944,Bondi}. This process has been studied analytically and numerically in both Newtonian and relativistic regimes for example in \citep{frixell1987,matsuda1987,shima1988,sawada1989,matsuda1992,matsuda1991} and in \citep{petrich1988,Font1998a,donmez,cruz2012,Lora_Clavijo_2013} respectively.

The BHL accretion by itself becomes more realistic as more ingredients are added up to the wind and accretor models. Recent versions of BHL analyses include the relativistic accretion a supersonic fluid onto a spinning black hole in 3D \citep{ours2015}, the study of 2.5D Hydrodynamic BHL accretion onto black holes including magnetic fields \citep{penner2010,takahashi}, the BHL accretion onto black holes including and the coupling between radiation and fluid \citep{zannoti2011,park2013}. 
In the Newtonian regime the study of magnetized BHL accretion onto a neutron star can be found in \citep{toropina}, also in \citep{AALee} the process of a magnetized plasma onto black hole was studied using a Newtonian model of black hole. 

Astrophysically motivated scenarios for the BHL accretion process involve the  properties of the gas around the black hole, including its equation of state, possible radiation transport processes that eventually could affect the dynamics of the plasma and magnetic field configurations. The velocity of the wind is another important parameter, directly related to the possible cause of the motion of the black hole. For instance, it has been found using numerical simulations, that the collision of two black holes with appropriate initial spins and masses may produce final black holes that travel at very high speeds, including velocities for supermassive black holes of the order of 100-1000 km/s in early analyses \citep{Gonzalez}, and up to 15000 km/s  \citep{kick} in more recent studies. Astronomical observations indicate that a candidate to be a wandering black hole, whose velocity has been modeled with this process,  the object known as QSO 3C 186 \citep{chiaberge2017}, traveling at a speed of 2100km/s, which could be the result of the collision of two black holes with appropriate initial orbital and spin parameters \citep{Lousto2017}. Mechanisms that promote stellar mass black holes to move in the interstellar space are the supernovae natal kicks, for example as found in \citep{https://doi.org/10.48550/arxiv.2201.13296}, which have considerable lower velocities. Even though it is possible to carry out simulations of the BHL accretion process with these low speeds (e. g. \cite{GonzalezGuzman2018}), technically such simulations require a considerable amount of resources due to the big numerical domain required for the accretion regime to hold, where the spatial scale can be hundreds of horizon radii for a supersonic scenario. In our analysis we use rather high values of the velocity, which allows the use of a smaller numerical domain, with appropriate numerical parameters for accurate simulations that suffice to illustrate the effects of the magnetic field on the wind in a spatial scale of a few horizon radii. Even in such case, also very high velocity realistic scenarios exist, for example the matter model using BHL accretion on black holes prior to mergers in GW sources within the common envelope stage \citep{Cruz_Osorio_2020}.

Among the astrophysical applications of the BHL accretion we find recent advances, that include the application of the shock-cone flip-flop instability \citep{donmez} and the shock cone vibrations  \citep{Lora_Clavijo_2013} that occur during the BHL accretion, as models of X-ray quasi periodic oscillators (QPOs); the BHL has been studied in environments with non-trivial density gradients  \citep{Lora_Clavijo_2015} and in the presence of small rigid bodies \citep{Cruz_Osorio_2017}; BHL has been also studied in binary stars \citep{binarystars}, and has also been proposed as a possible ignition mechanism of type Ia supernovae \citep{Emilio}; very recently the formation of jets in BHL processes has been also presented \citep{https://doi.org/10.48550/arxiv.2201.11753}, as well as the influence of BHL in sources of gravitational waves \citep{Cruz_Osorio_2020}.

The degree of applicability of models with more ingredients would depend on the observational resolution of black hole horizon size scale, for example using the Event Horizon Telescope array, which has revealed high resolution images of plasma surrounding the supermassive black hole at the centre of M87 \citep{EHTM871} and at the center of the Milky Way's Sgr A* \citep{EHTSgrA}. Other scenarios like the interesting moving black hole associated to the quasar 3C 186  \citep{chiaberge2017}, that could be a kicked black hole moving through the galaxy medium resulting from the merger of two black holes  \citep{Lousto2017}, would require a resolution currently out of reach due to the distance from earth, although resolution is expected to always improve.

In order to contribute to the addition of ingredients modeling the BHL process onto a spinning black hole, in this paper we study the 3D supersonic accretion of magnetized winds within the ideal magnetohydrodynamics approximation (MHD) and compare the general results with the accretion of a purely Hydrodynamical gas (HD). In this sense, the present paper is a follow up of \citep{ours2015}. We study the evolution of the MHD variables and describe the differences between the accretion of a purely hydrodynamical fluid and a magnetized plasma. In our analysis we assume the black hole and wind to be initially immersed in a constant magnetic field,  aligned with the direction of the axis of rotation of the black hole. 
In order to investigate the potential properties of a general case scenario, we choose three principal wind directions with respect  to the spin of the black hole: wind parallel to the axis of rotation, diagonal wind and a wind perpendicular to the axis of rotation of the black hole. Notice that the last two cases can only be studied in full 3D without symmetries.

The paper is organized as follows. In section \ref{sec:numerical_set_up} we describe the ideal MHD equations modeling the magnetized fluid and the numerical methods used. In section \ref{sec:results} we present the set of configurations we experiment with and the main aspects we compare between the HD and MHD scenarios. In \ref{sec:conclusions} we describe the conclusions from our analysis and in the appendix we show convergence tests of our simulations.

% --------------------------------------------
% ---------->     SECTION     <----------
% --------------------------------------------
\section{The wind model}
\label{sec:numerical_set_up}

% ---------->     Subsection     <----------
\subsection{Equations and numerical methods}

The plasma is modeled with a magnetized fluid that obeys the ideal MHD, which assumes infinite electric conductivity and the electric field measured by a comoving observer set to zero. The stress-energy tensor of such fluid is explicitly

\begin{equation}
{ T }^{ \mu \nu  }=(\rho h+{ b }^{ 2 }){ u }^{ \mu  }{ u }^{ \nu  }+\left( p+\frac { { b }^{ 2 } }{ 2 }  \right)g^{\mu\nu}-{ b }^{ \mu  }{ b }^{ \nu  },
\end{equation}

\noindent where $\rho$ is the rest-mass density, $p$ the pressure, $b^\mu$ the magnetic field measured by a comoving observer, ${ u }^{ \mu  }$ is the 4-velocity, $h \equiv 1 + \epsilon + p/\rho$ the specific enthalpy, $\epsilon$ the specific internal energy and $g^{\mu\nu}$ are the contravariant components of the 4-metric.

The equations for this matter field are those of the general relativistic magnetohydrodynamics (GRMHD). These can be  written as a flux conservative system that assumes a standard 3+1 decomposition of the space-time. The space-time metric described in Cartesian coordinates $(t,x^i)$ is given by $ds^2 = (-\alpha^2 + \beta^i \beta_i)dt^2+2\beta_i dt dx^i + \gamma_{ij}dx^i dx^j$, where $\alpha$ is the lapse function and $\beta^{i}$ the components of the shift vector associated to the 3+1 decomposition of the space-time, and $\gamma_{ij}$ are the components of the 3-metric of spatial hypersurfaces used to foliate the space-time. In these terms, the GRMHD equations according to the Valencia formulation are written as \citep{banyuls1997}:

\begin{equation}
  \label{eq:conservative}
  \partial_t {\bf u} + \partial_{x^i} {\bf F}^{(i)} ({\bf u}) = {\bf
    S} ({\bf u}),
\end{equation}

\noindent  where the vector
${\bf u}=\{D,S_i,\tau, B^k\}$ contains the following conserved variables, $D$ the generalized rest mass density of the fluid, $S_i$  the momentum components along in each direction, $\tau$  the internal energy, $B^k$ the magnetic field measured by an eulerian observer,  ${\bf F}^{(i)} ({\bf u})$ the fluxes and ${\bf S} ({\bf u})$ a sources vector. In terms of the primitive variables of the fluid elements, the conserved variables are defined by

\begin{eqnarray}
  \label{eq:prim2con}
   D &=& \sqrt{\gamma}\rho W \nonumber \\
   S_i &=& \sqrt{\gamma}[ (\rho h+{ b }^{ 2 }){ W }^{ 2 }{ v }_{ j }-{ \alpha  }{ b }^{ 0 }{ b }_{ j } ]\nonumber\\
   \tau &=& \sqrt{\gamma}[(\rho h{ +{ b }^{ 2 })W }^{ 2 }-\left( p+\frac { { b }^{ 2 } }{ 2 }  \right) -{ \alpha  }^{ 2 }({ b }^{ 0 })^{ 2 }-\rho W ]\nonumber\\
   B^i &= & \sqrt{\gamma}W\left[{ b }^{ i }-\alpha { b }^{ 0 }\left( { v }^{ i }-\frac { { \beta  }^{ i } }{ \alpha  }  \right)\right]\nonumber, 
\end{eqnarray}

\noindent where $\gamma$ is the determinant of the spatial 3-metric $\gamma_{ij}$ of the three-dimensional spatial slices which foliate the space-time, $W=(1-\gamma_{ij}v^i v^j)^{-1/2}$ is the Lorentz factor and ${ b }^{ 0 }=WB^{ i }{ v }_{ i }/\alpha $. As usual, the system of equations is closed with an equation of state specified below. In terms of primitive and conservative variables the fluxes and sources are

{\small
\begin{eqnarray}
        \  {\mathbf F}^i({\bf u}) &=& \left(
	      \begin{array}{c}
			(\alpha v^i - \beta^i)D\\
	        (\alpha{ v }^{ i }-{ \beta  }^{ i }){ S }_{ j } +\alpha\sqrt{\gamma}\left( p+\frac { { b }^{ 2 } }{ 2 }  \right) { \delta  }_{ j }^{ i }-\alpha\sqrt{\gamma}{ b }_{ j }{ B }^{ i }/W\\
	         \left( \alpha{ v }^{ i }-{ \beta  }^{ i }   \right)\tau +\alpha\sqrt{\gamma}{ \left( p+\frac { { b }^{ 2 } }{ 2 }  \right)  }v^i -\alpha^2\sqrt{\gamma} { b }^{ 0 }{ B }^{ i }/W \\
			 \left(\alpha { v }^{ i }-{ \beta  }^{ i }\right) { B }^{ k }-\left( \alpha{ v }^{ k }-{ \beta  }^{ k }  \right) { B }^{ i }
	      \end{array} \right),\nonumber\\
        \  \mathbf{S({\bf u})} &=& \left(
	      \begin{array}{c}
			0\\
	        T^{\mu\nu}\big (\partial_\mu g_{\nu j} +\Gamma^\delta_{\mu\nu} g_{\delta j}\big )\\
	        \alpha \big (T^{\mu0}\partial_\mu \ln \alpha - T^{\mu\nu}\Gamma^0_{\mu\nu} \big) \\
			0
	      \end{array} \right),\label{eq:fluxes}
\end{eqnarray}
}

\noindent where $g_{\mu\nu}$ are the covariant components of the 4-metric and $\Gamma^{\delta}{}_{\mu\nu}$ the Christoffel symbols of the space-time.
We solve the system of equations (\ref{eq:conservative}-\ref{eq:fluxes}) using the publically available GRHydro thorn \citep{baiotti2005}, within the Cactus Einstein Toolkit (ETK) code  \citep{ETK2012}.  We use the high resolution shock capturing methods provided to solve the GRMHD equations. Specifically our simulations use the HLLE numerical flux formula and the minmod reconstructor. In order to preserve the magnetic field divergence near to zero, we use the constraint transport method \citep{balsara} implemented within the GRHydro thorn. For the integration in time we use a fourth order Runge-Kutta method. What we added to the ETK is a module that applies appropriate boundary conditions in the upstream boundary, which is the part of the boundary from which we inject the wind into the domain, where we set the density and velocity field to their initial values during the evolution. On the other hand we implement out-flux  boundary conditions in the downstream boundary, which is the part of the boundary through which the wind is expected to leave the domain. These conditions  are a key ingredient in the accretion of winds in numerical domains that contain the accretion sphere. In order to avoid divergences of the  variables, the GRHydro thorn implements an atmosphere that is triggered during the primitive variables calculation for tiny or negative values of the density. For that we use a floor density value of $10^{-12}$ in code units, and then the pressure and internal energy are set to consistent values. In our results the density never approaches such small values, including the cases where rarefaction spots are formed.

% ----->     SUBSECTION     <-----
\subsection{Description of space-time and wind}

We describe the space-time metric of the black hole of mass $M$ and spin ${\bf S}=a\hat{z}$ using Kerr-Schild (KS) horizon penetrating coordinates:

\begin{eqnarray}
ds^2 &=& \left(\eta_{\mu\nu} + \frac{2Mr^3}{r^4+a^2z^2}l_{\mu}l_{\nu}\right)dx^{\mu}dx^{\nu},\nonumber\\
l_{\mu} &=& \left(1,\frac{rx+ay}{r^2+a^2},\frac{ry-ax}{r^2+a^2},\frac{z}{r}\right),\nonumber\\
r &=& \sqrt{\frac{r_*^{2}-a^2+\sqrt{(r_*^{2}-a^2)^2+4a^2z^2}}{2}},\nonumber\\
r_{*} &=& \sqrt{x^2 + y^2 + z^2},
\end{eqnarray}

\noindent where $\eta_{\mu\nu} $ is the flat metric.

The {\it wind}  is set initially as a spatially constant rest mass density ideal gas $\rho = 1\times10^{-6}\;[1/M^2]$, moving toward the black hole at a given asymptotic supersonic velocity $v_{\infty}^2=v^iv_i$. We assume the fluid obeys a gamma-law equation of state $p=(\Gamma-1)\rho \epsilon$, and consider a relativistic fluid with adiabatic index $\Gamma=4/3$. We use the asymptotic speed of sound $c_{s \infty}=0.05$ in order to have a sufficiently slow wind but still supersonic  and to compare with previous hydrodynamical results. The initial fluid pressure is written as $p_{\mathrm{ini}} = c_{\mathrm{s} \infty}^2 \rho_{\mathrm{ini}}/(\Gamma - c_{\mathrm{s} \infty}^2 \Gamma_1)$, where $\Gamma_1=\Gamma/(\Gamma -1 )$. In order to avoid negative and zero values of the pressure we choose the sound speed such that $c_{\mathrm{s}\infty} < \sqrt{\Gamma - 1}$. Finally, the initial  specific internal energy $\epsilon$ is reconstructed using the equation of state.

The magnetic field is defined initially to be constant and parallel to the spin of the black hole ${\bf B}=B_0\hat{z}$. We choose two representative values of magnetic field strength $B_{0,strong}=1\times10^{-5}\;[1/M]$ and $B_{0,weak}=1\times10^{-10}\;[1/M]$. After the initial time the magnetic field evolves according to the GRMHD equations and responds to the evolution of the plasma.

We focus on three representative scenarios, in which the wind is assumed to have different direction  with respect to the black hole spin ${\bf S}$. These three different orientations of the wind are represented by $\uparrow~ \leftarrow ~\nearrow$ in our tables and correspond to directions $\hat{z}$, $\hat{x}$ and $\hat{x}+\hat{z}$ respectively. The black hole spin is denoted by $\Uparrow$.

We also use the excision method inside the black hole horizon in order to avoid the variables to interact with the black hole's singularity \citep{hawke2005}. We performed all of our evolution runs using an isotropic cubic grid $\Delta x = \Delta y = \Delta z$  with base resolution $\Delta x=0.5M$ and one refinement level with resolution $\Delta x =0.25M$. The base domain is set to $[-30M,30M]^3$ and is approximately twice as big as a sphere of the accretion radius $r_{acc}=M/(c_{s \infty}^{2}+v_{\infty}^2)$. This is important because in case the domain is smaller than a sphere with radius equal to the accretion radius, the flow could enter the wind regime. 

Continuing with the set up of the initial configuration, another important property is the size scale. This is usually set by the accretion radius defined by $r_{acc}=r_{hor}/(v_{\infty}^{2} + c_{s\infty}^2)$, where $r_{hor}$ is the radius of an accretor, in our case the horizon radius of the black hole. The accretion radius has the information of how fast or slow the wind is and it is important because as studied in \citep{foglizzo2005},  the parameter that could trigger a possible flip-flop instability in the Bondi-Hoyle process, is the relative size of the accretor with respect to the accretion radius $r_{hor}/r_{acc} = v_{\infty}^{2} + c_{s\infty}^2$. Sometimes this is called the accretor size \citep{foglizzo2005}, but it is actually a size relative to the accretion radius that indicates how fast or slow the wind is. In our study we use two wind velocities corresponding to a slow case $r_{hor}/r_{acc} = 0.065$ and a fast case $r_{hor}/r_{acc} = 0.25$ with asymptotic velocity $v_{\infty}=0.25c$ and $0.5c$ respectively. 

Finally, we set the dimensionless spin of the black hole to ${\bf S}=0.8 \hat{z}$, which is comparable to the one estimated by numerical simulations of the QSO 3C186 quasar's kicked black hole \citep{Lousto2017}.

% --------------------------------------------
% ---------->     SECTION     <----------
% --------------------------------------------
\section{Results}
\label{sec:results}

\begin{figure}
\centering
\includegraphics[width=4.15cm]{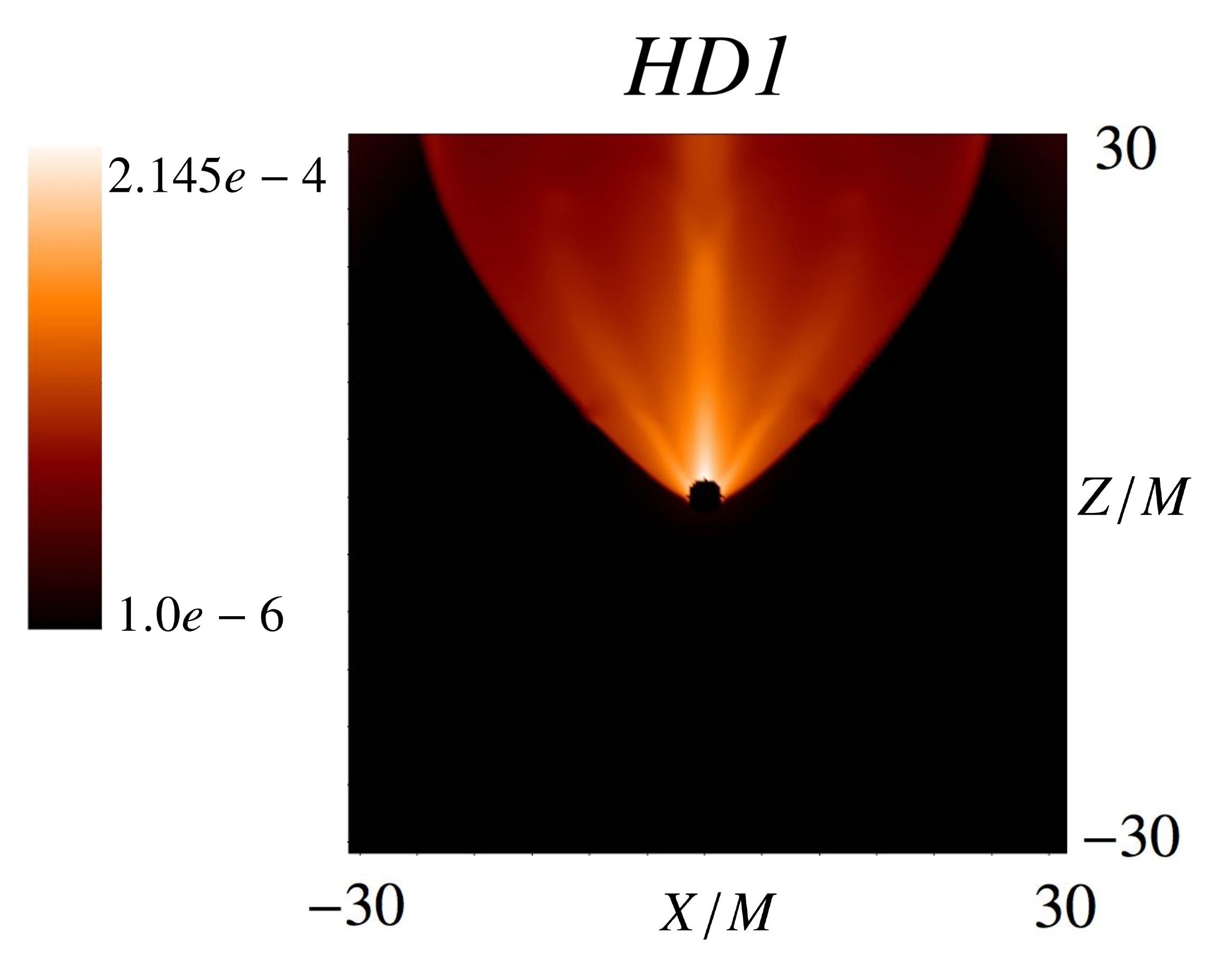}
\includegraphics[width=4.15cm]{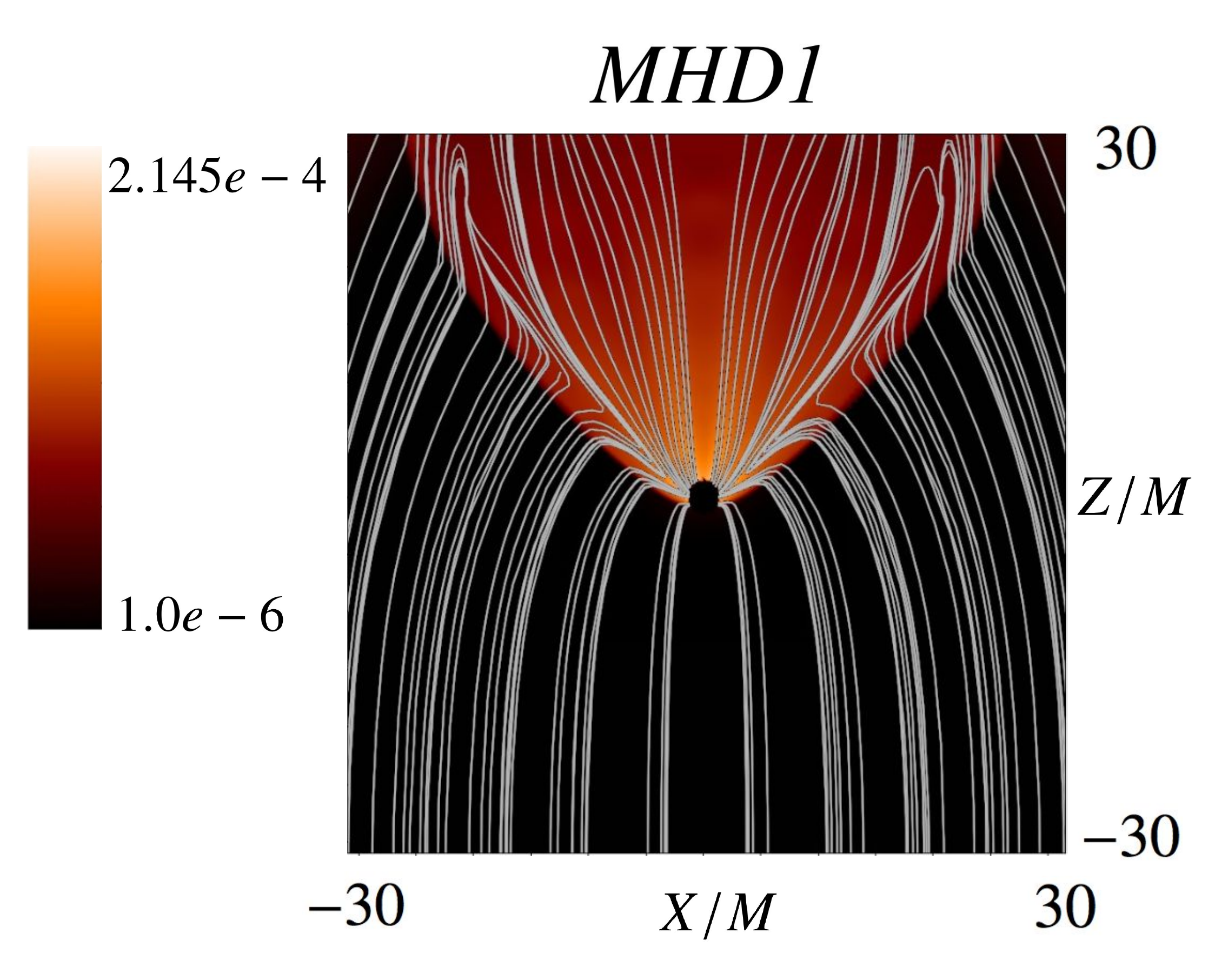}
\includegraphics[width=4.15cm]{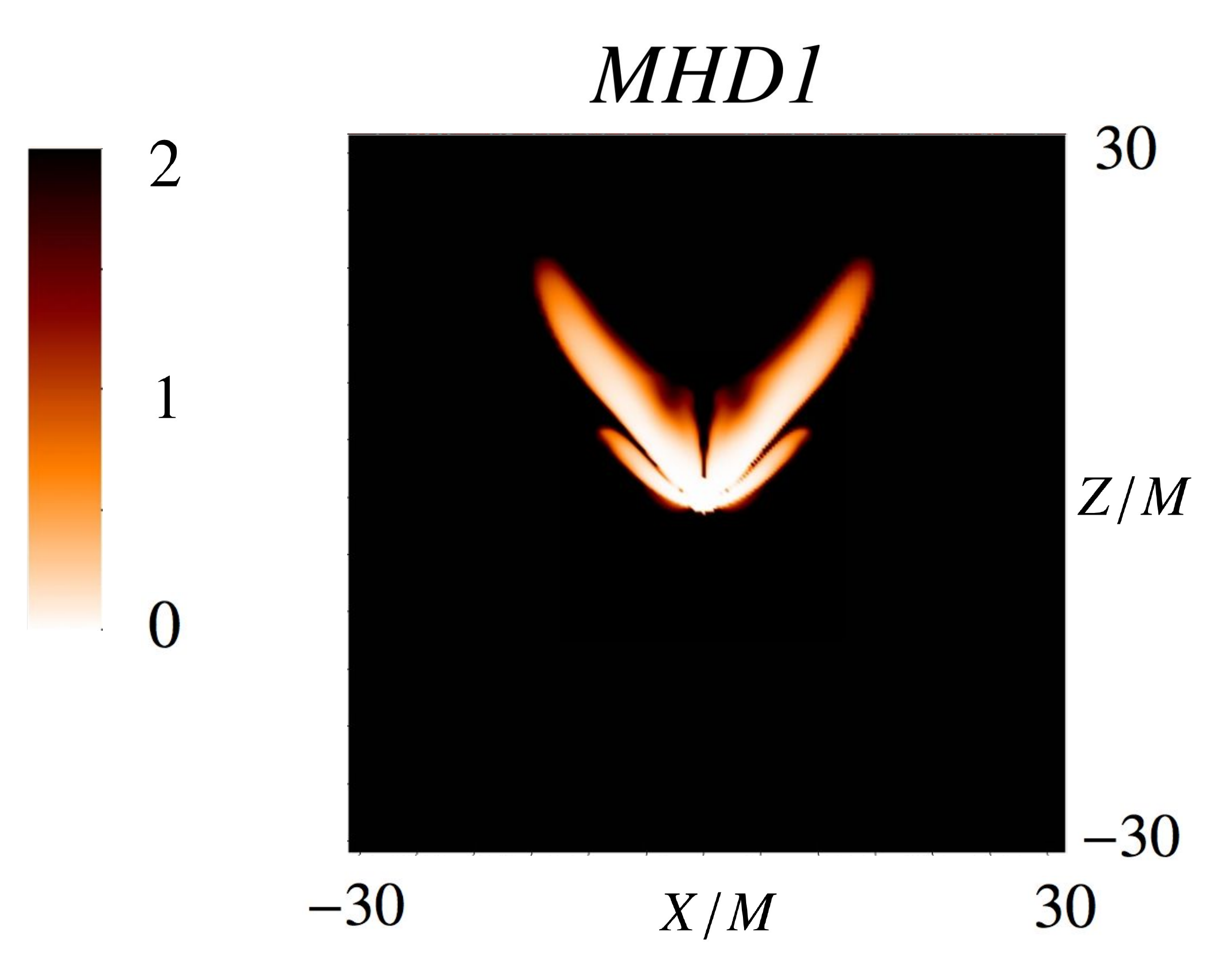}
\includegraphics[width=4.15cm]{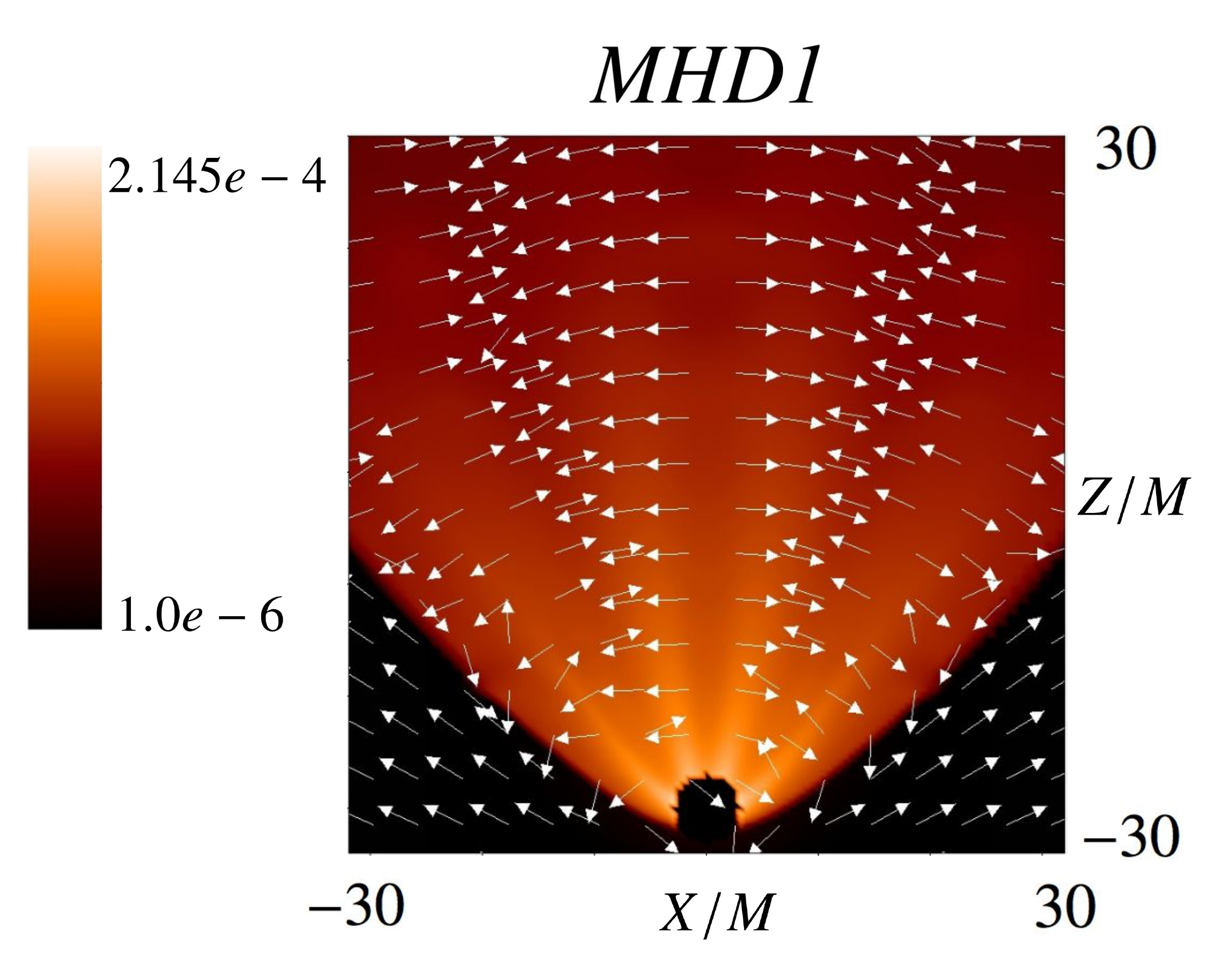}
\caption{\label{fig:northslow} Snapshot on the $y=0$ plane of the density at time $t=1000M$ when the accretion is already stationary, for  cases HD1 and MHD1 models $\uparrow \Uparrow$ and $r_{hor}/r_{acc}=0.065$ shown in the top. We show the magnetic field lines in the MHD1 case indicating how they bend toward the cone and after that continue to their asymptotic vertical direction. In the bottom we show the $\beta$ parameter and the shock cone superposed with a field of vectors indicating the Lorentz force direction. The shock cone in the presence of magnetic field is slightly wider than in the purely HD case and in the MHD case the maximum of the gas density density ($1.3\times 10^{-4}$) is  smaller than the HD case ($2.1\times 10^{-4}$).}
\end{figure}	

\begin{figure}
\centering
\includegraphics[width=4.15cm]{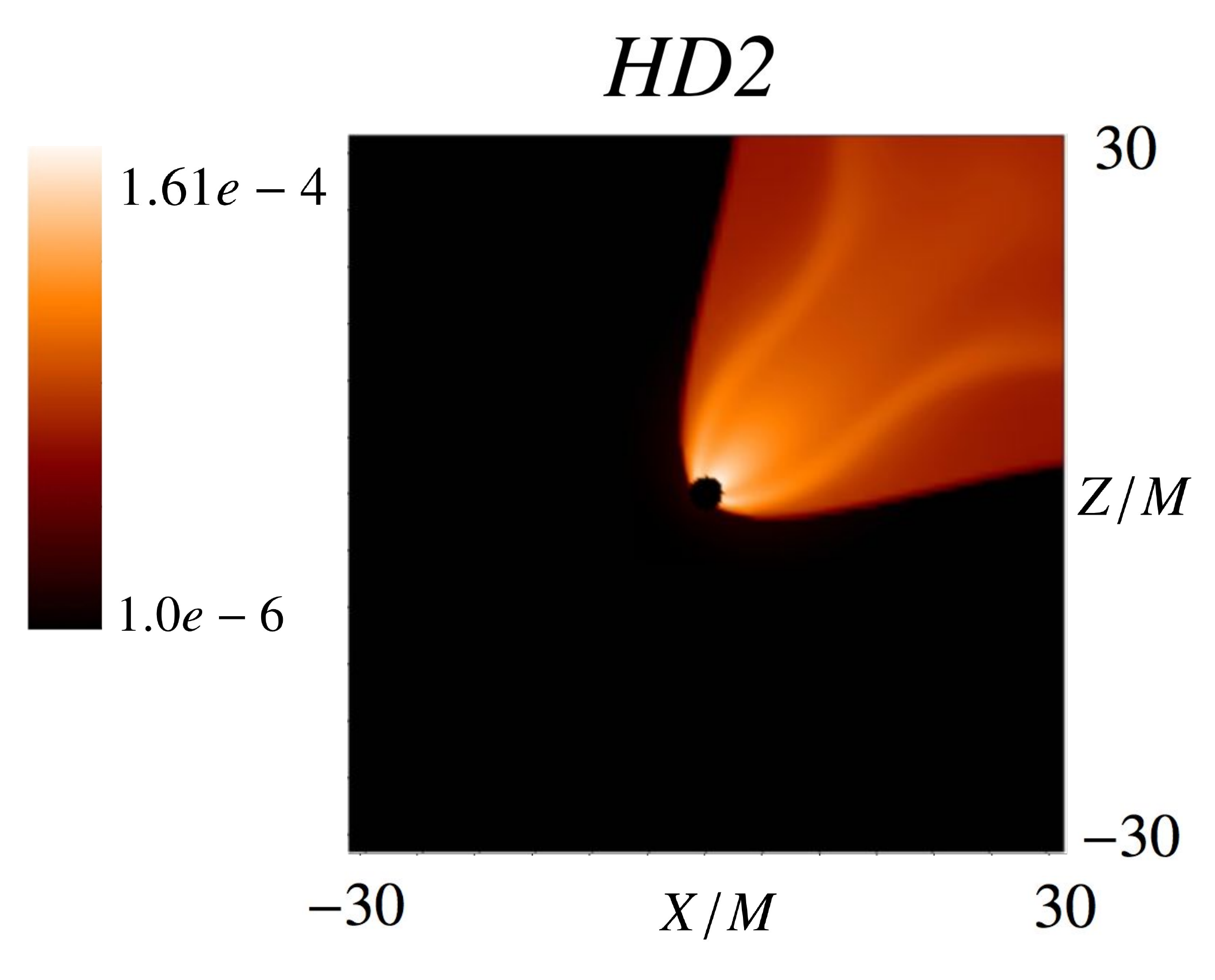}
\includegraphics[width=4.15cm]{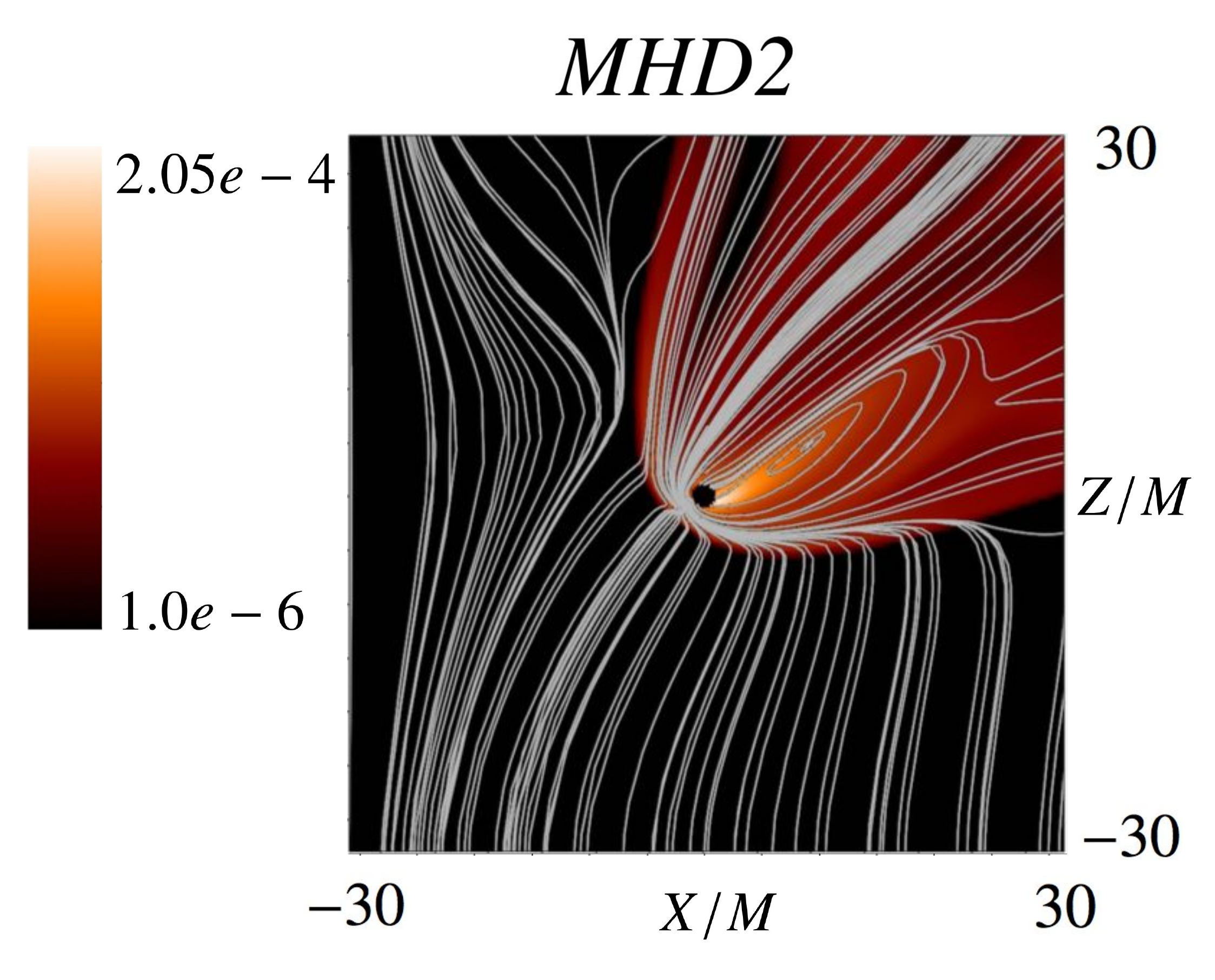}\\
\includegraphics[width=4.15cm]{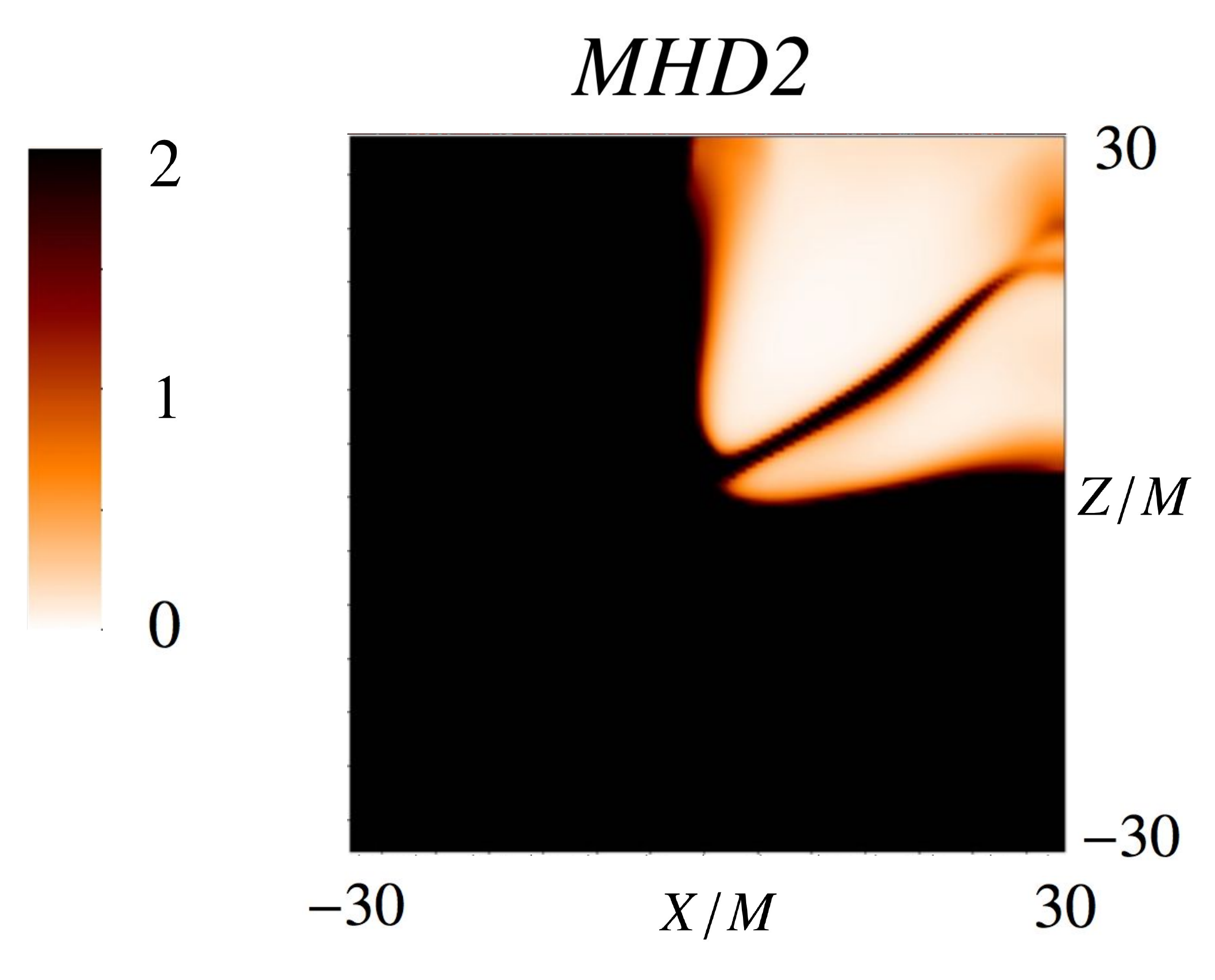}
\includegraphics[width=4.15cm]{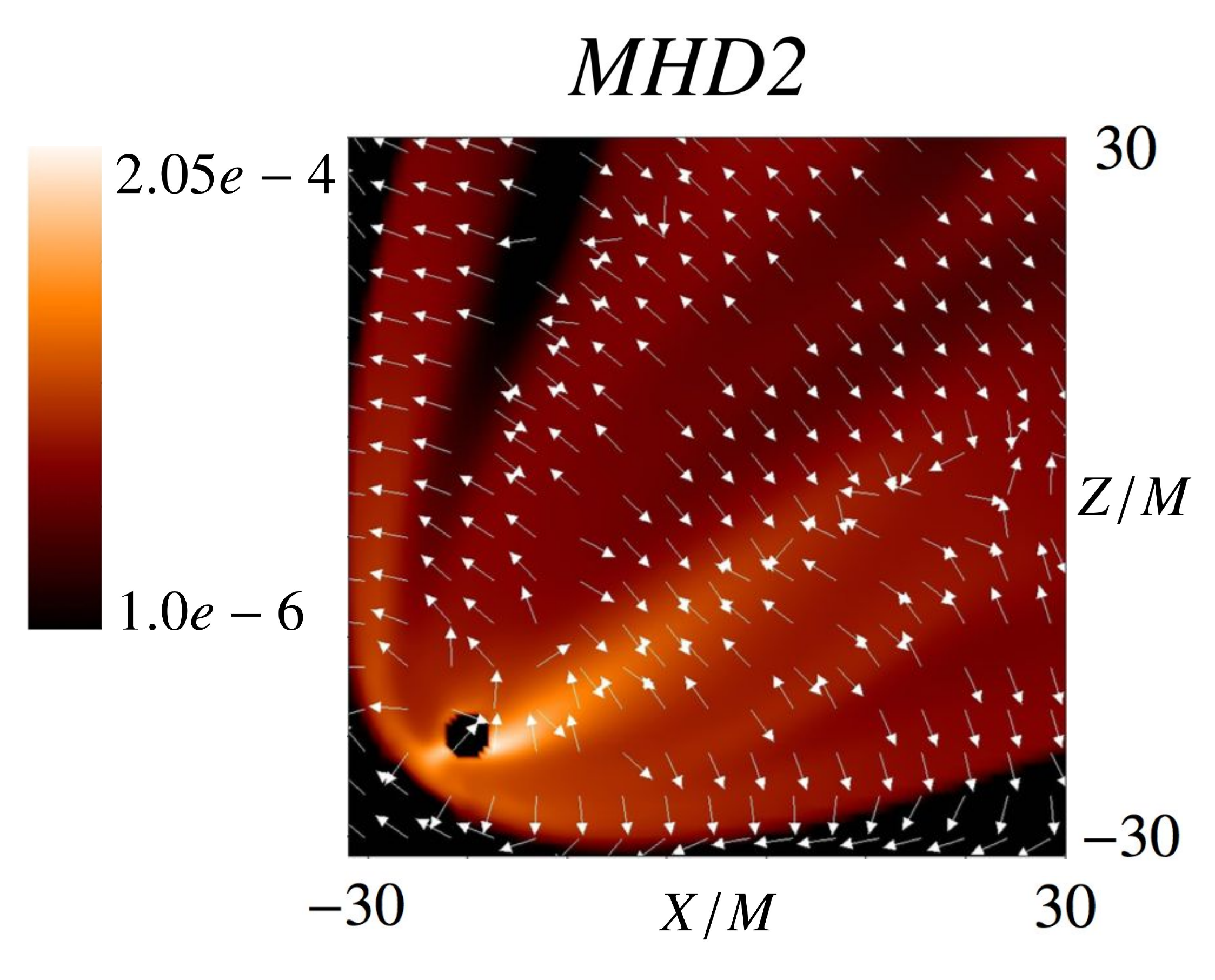}
\caption{\label{fig:diagonal} In the top we show a snapshot on the $y=0$ plane of the density at  $t=1000M$, for the HD2 and MHD2 cases with orientation $\nearrow \Uparrow $ and $r_{hor}/r_{acc}=0.065$. The first important difference between the HD and MHD is that in the later, the shock-cone is detached from the black hole, that is, the black hole is contained within the high density zone. The second is that the shock cone in the presence of the magnetic field is wider than in the pure HD case. Finally the third one is that in the MHD case there is a low density region.
In the  MHD2 case we superpose the magnetic field lines, which show asymptotically the vertical direction, however shows a complex structure at the boundary of the shock cone. The structure of field lines would be interesting to see in detail within the context of resistive MHD, because there are some candidates to X-points that eventually could trigger magnetic reconnection. 
In the bottom we show a plot of the $\beta$ parameter, which shows that the magnetic field dominates in the region of the shock-cone. In the final plot we superpose the density of the plasma with a field of arrows indicating the direction of the Lorentz force. It can be seen that the arrows precisely indicate that the Lorentz force pulls the plasma out of the rarified zone.}
\end{figure}

With the above set of physical parameters, including two wind velocities, two magnetic filed strengths, we performed a series of simulations for the wind orientations summarized in Table \ref{tab:tabla1}, where we include the purely Hydrodynamical counterparts in order to compare the impact of the magnetic field on the process. 
The simulations start with the constant values of the wind variables except in the excision region. During a transient stage the fluid and magnetic field interact until they approach nearly stationary configurations, where the bow shock if any and the shock cone are formed, and in the MHD cases the magnetic field also stabilizes. This stationary stage is the one we illustrate in the results discussed below.

\begin{figure}
\centering
\includegraphics[width=4.15cm]{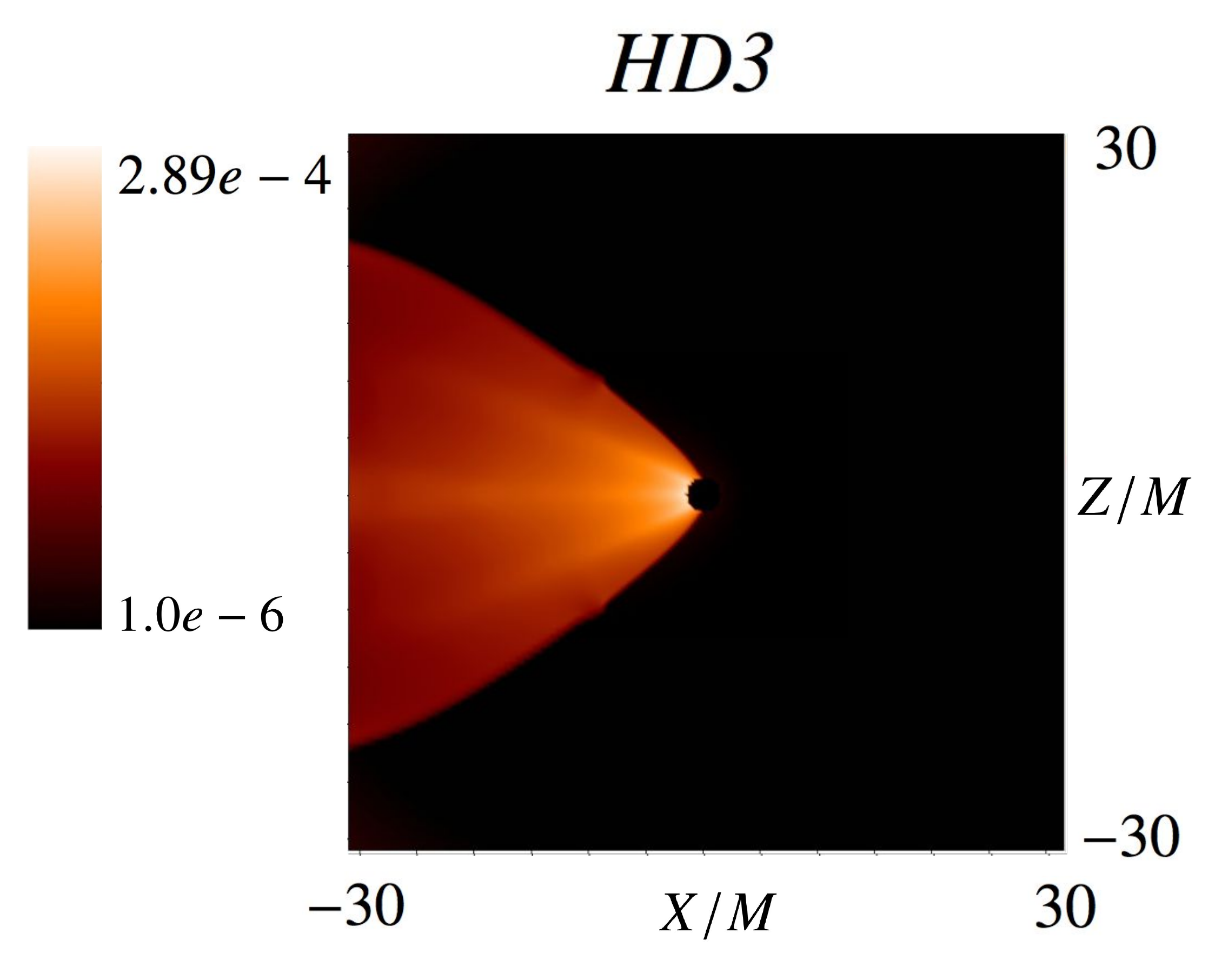}
\includegraphics[width=4.15cm]{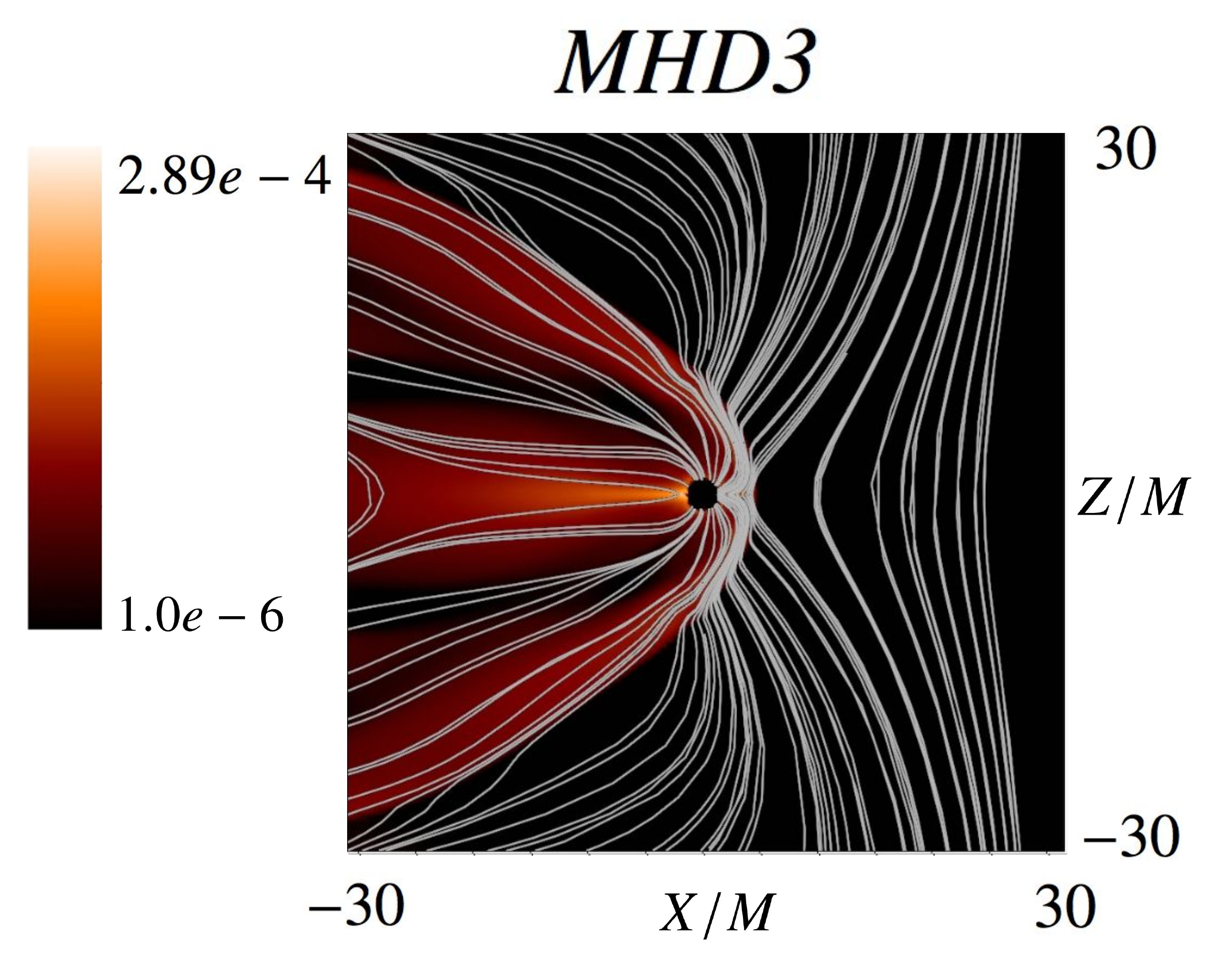}\\
\includegraphics[width=4.15cm]{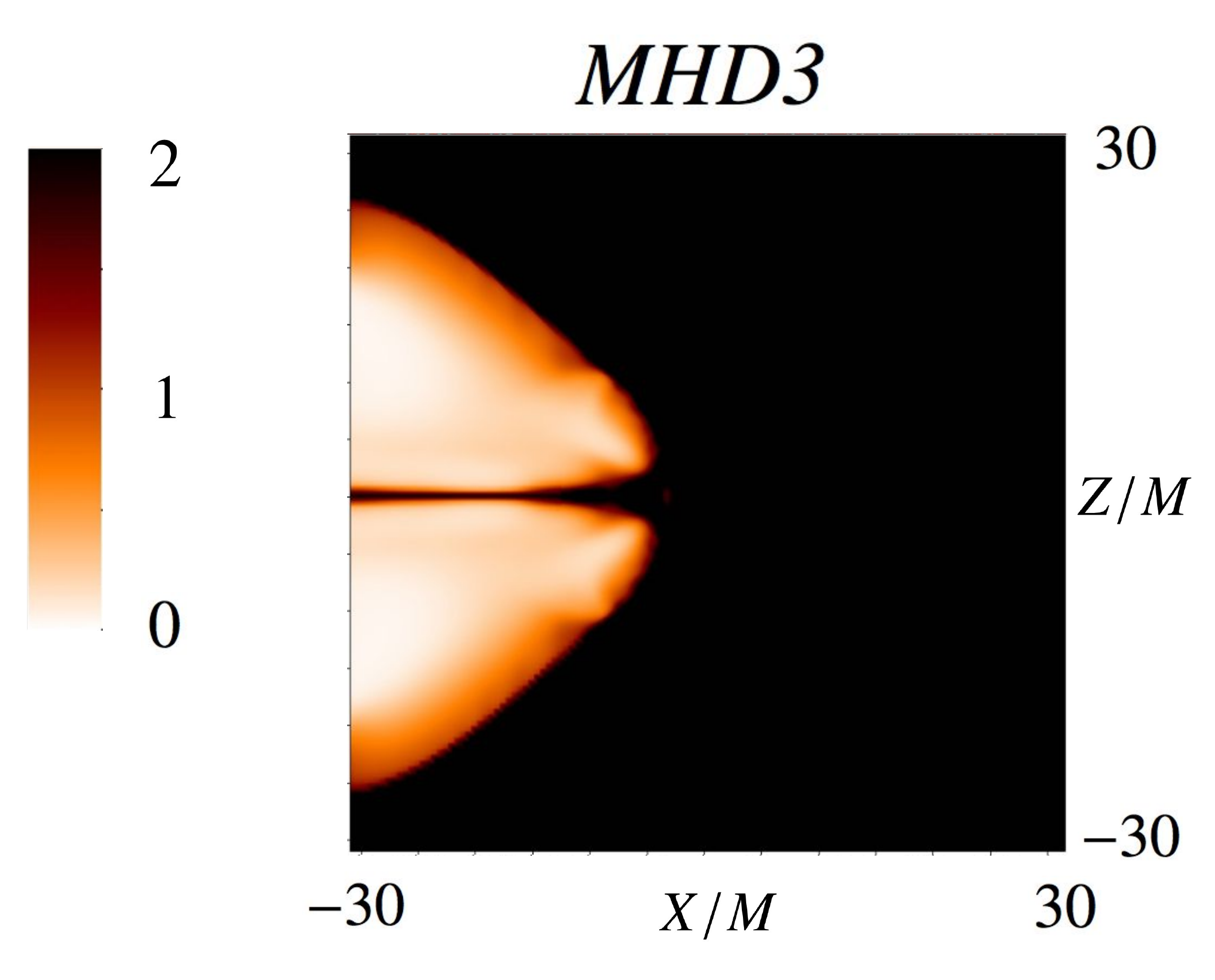}
\includegraphics[width=4.15cm]{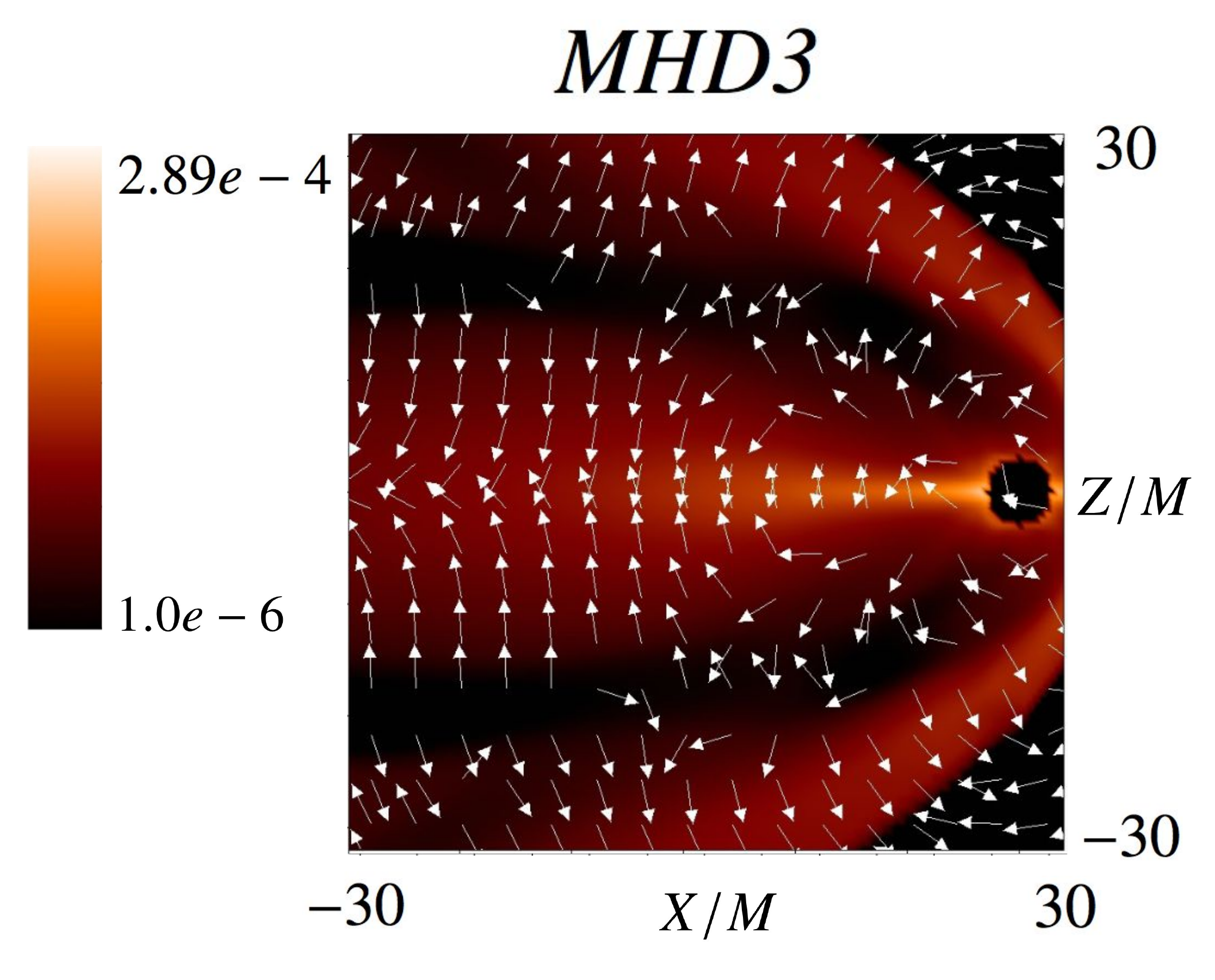}
\caption{\label{fig:west} As in the previous case, we show in the top a snapshot of the rest mass density on the $y=0$ plane at time $t=1000M$ for the HD3 and MHD3 models with $r_{hor}/r_{acc}=0.065$ for the case $ \Uparrow\leftarrow$. Again a difference is that in the MHD case the shock-cone is  detached from the black hole and the magnetic field lines shows  an interesting distortion near the bow-shock. Due to the bitant symmetry of this case, there are two symmetric rarified regions.  In the bottom we show that the region $\beta<1$ appears again within the shock-cone region. In the bottom-right panel we show that the Lorentz force points in the direction in which the plasma should move to produce the low density spots.}
\end{figure}

\begin{figure}
\centering
\includegraphics[width=4.15cm]{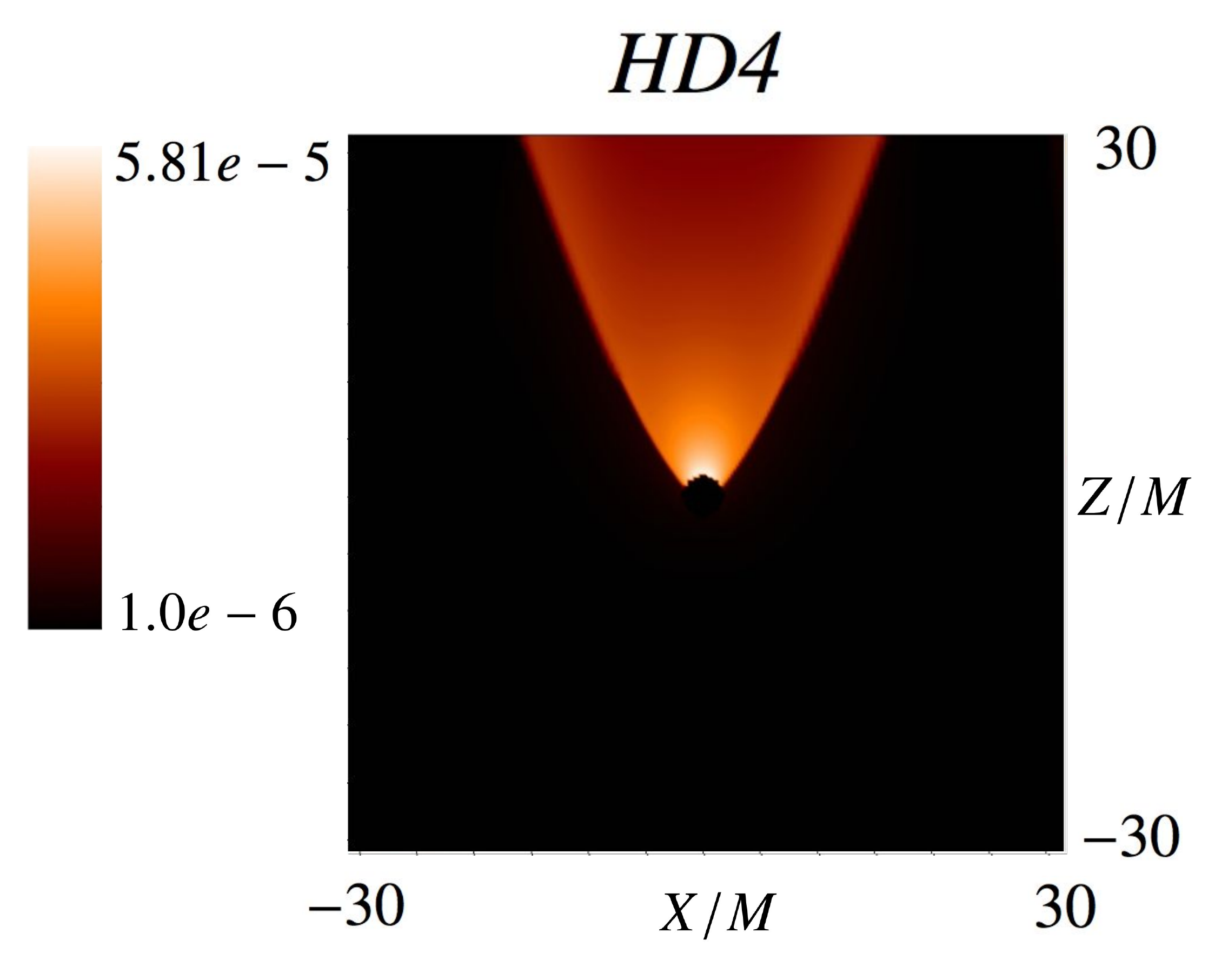}
\includegraphics[width=4.15cm]{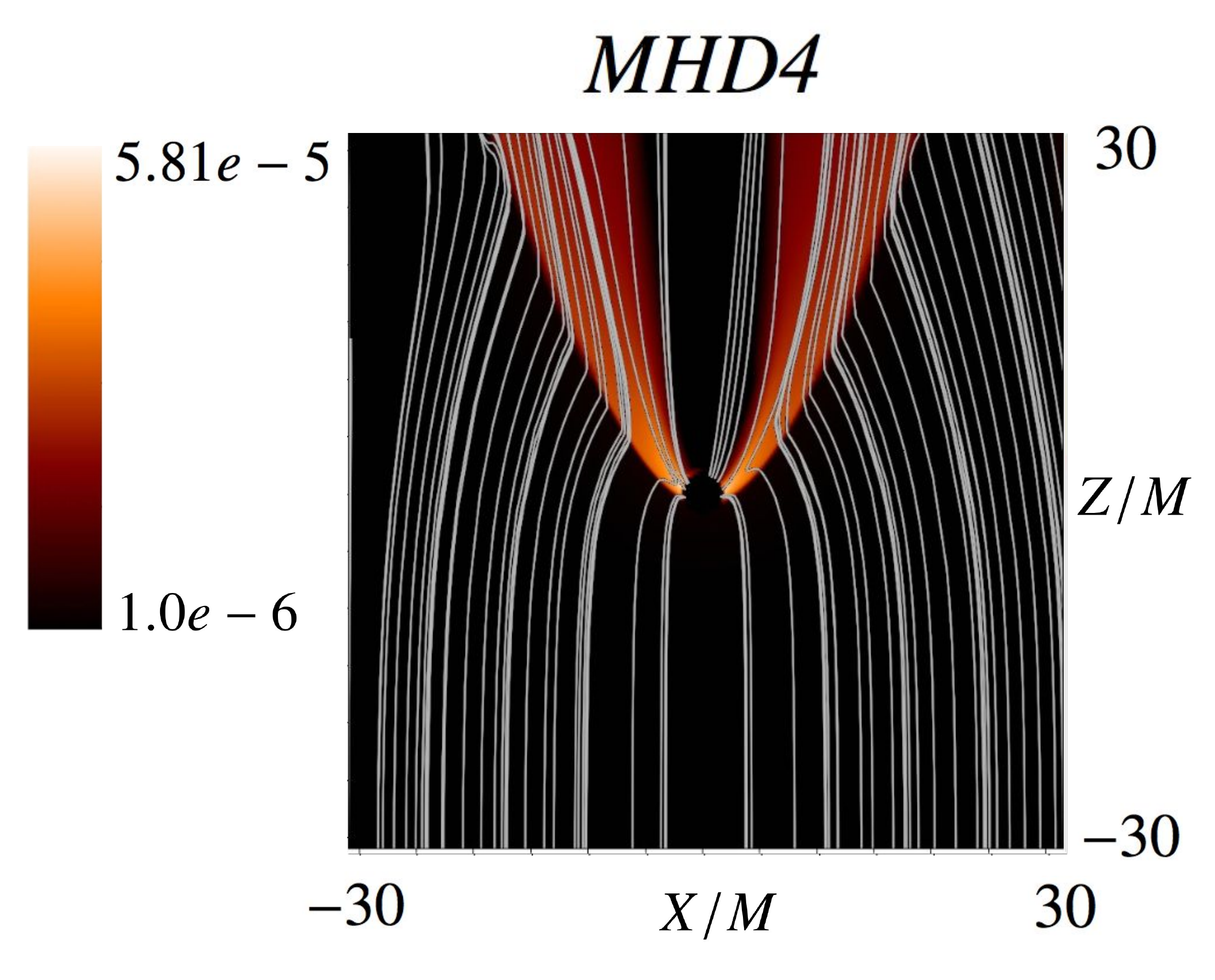}\\
\includegraphics[width=4.15cm]{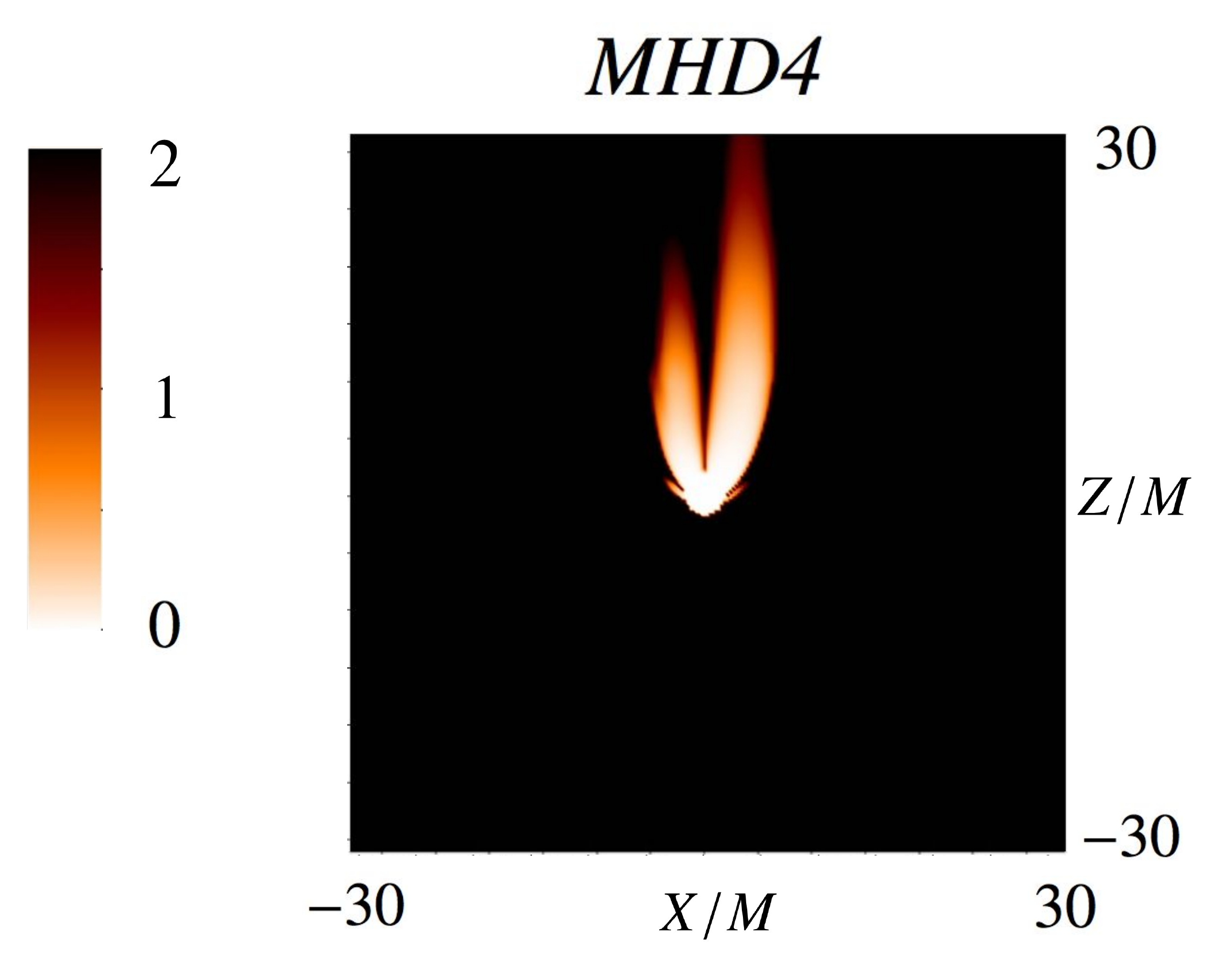}
\includegraphics[width=4.15cm]{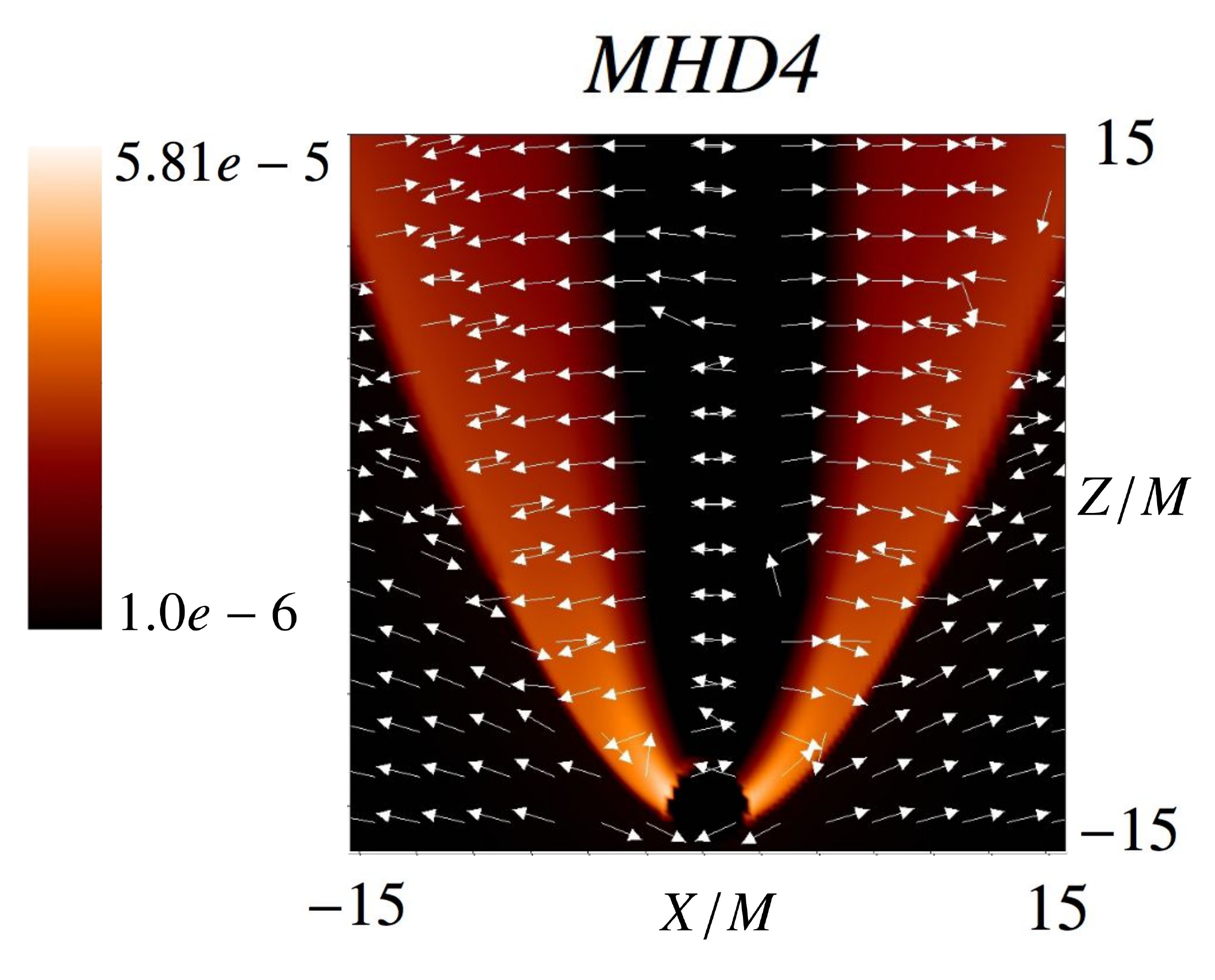}
\caption{\label{fig:northfast} This is the case of a fast wind. In the top we show a snapshot of the rest mass density on the $y=0$ plane at time $t=1000M$, for the HD4 and MHD4 models with $r_{hor}/r_{acc}=0.25$. Since this is a faster wind, in both HD and MHD cases, the shock cone is attached to the black hole. In the bottom we show the parameter $\beta$ which shows the magnetic field dominates in the shock-cone region. In this case there is a central rarified zone. According to the direction of the Lorentz force shown in the last panel, this force is the responsible for the depletion of the low density regions. The low velocity counterpart of the case is MHD1 in Figure \ref{fig:northslow}. The density in the rarified zone within the shock-cone of this MHD4 case is one order of magnitude smaller than in the MHD1 case. This is an example of how the velocity of the wind influences the properties of the system. }
\end{figure}

The general properties of the morphology and dynamics of the process once the evolution of the fluid has settled down to a nearly stationary regime are presented in Figures \ref{fig:northslow}, \ref{fig:diagonal}, \ref{fig:west} and \ref{fig:northfast},  in geometric units, corresponding to the four first models in Table \ref{tab:tabla1}. In these figures we show the rest mass density of the gas in the purely hydrodynamical scenario; the rest mass density of the plasma in the MHD cases with the magnetic field lines superposed, indicating the distortion due to the presence of the shock-cone, we remind that initially the magnetic field lines are parallel to the $z$ axis; we also show the value of $\beta=\frac{2p}{b_i b^i}$, which reveals that there is a region $\beta<1$ where the magnetic pressure dominates over the hydrodynamical pressure; finally we show the Lorentz force field, indicating the direction in which the plasma is being affected by this force.

{\it Stability.} Similarly to the purely hydrodynamical process, where no Flip-Flop (FF) kind of instability was found, even within the regime where it was predicted in the Newtonian theory \citep{ours2015}, in the case of  MHD there was not such instability either, even though some different  dynamical features were found.

The {\it bow shock}. Another important dependence of the orientation of the wind is the bow shock. In Fig. \ref{fig:diagonal} we show that for the MHD case the bow shock is detached from the black hole surface, a condition that usually triggers the shock cone instability in Newtonian systems \citep{foglizzo2005} but has been shown to be inoffensive in the relativistic case \citep{ours2015}.

{\it Shock cone angle.} As reported in \citep{penner2010} for the axisymmetric case, the open angle of the shock cone is bigger for the MHD than for the purely HD scenario. We confirm this result in the axisymmetric cases and also in the full non-symmetric diagonal $\nearrow \Uparrow$ and horizontal $ \Uparrow \leftarrow$ cases.

\begin{figure}
\centering
\includegraphics[width=8.5cm]{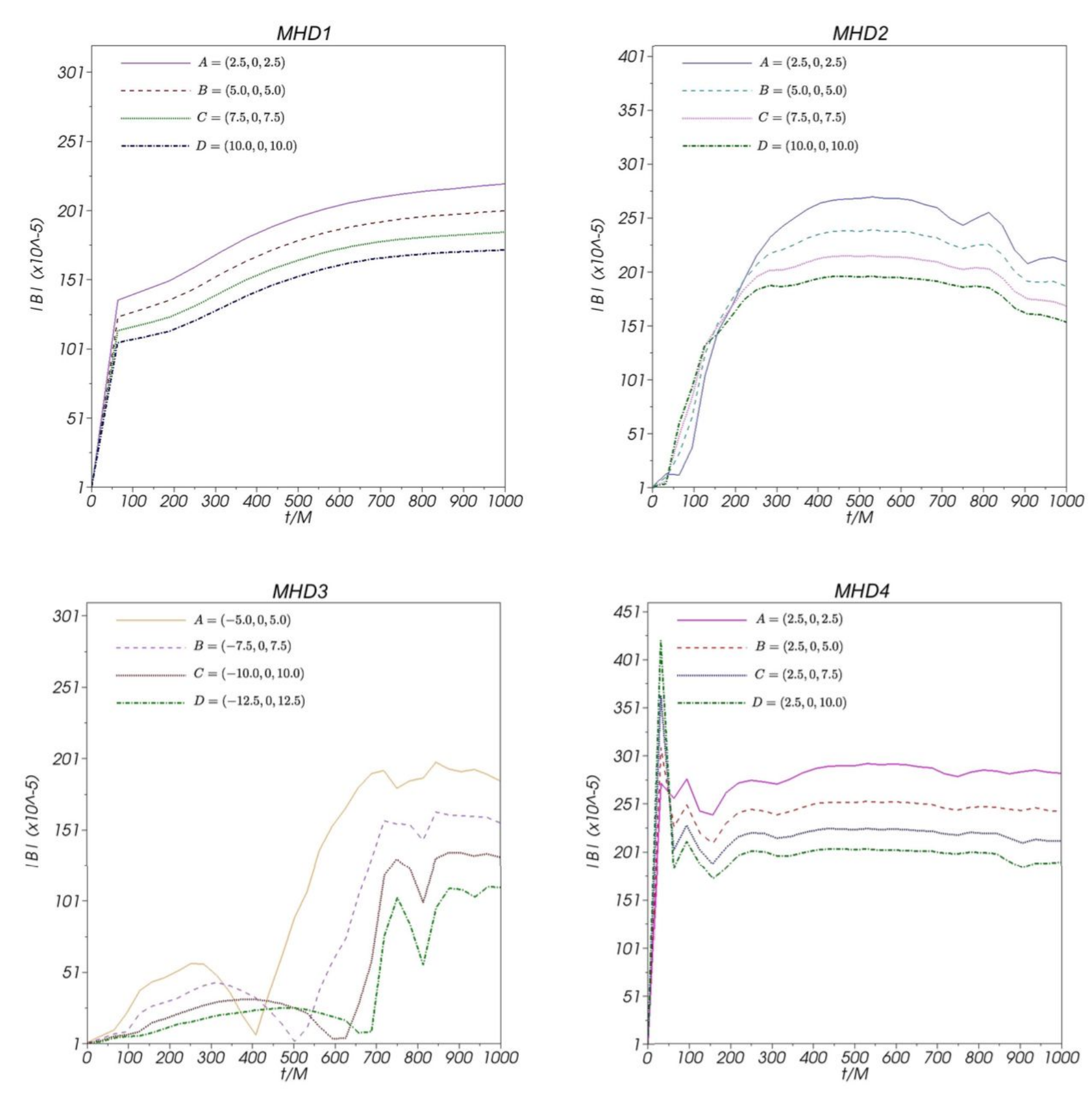}
\caption{\label{fig:ampli1} Time series of the magnetic field strength measured at four different points of the shock cone with a magnetic field $B_0=B_{0,strong}$ for models MHD1, MHD2, MHD3 and MHD4. For this we have chosen points located where the magnetic pressure dominates. These points can be identified in Fig. \ref{fig:northslow} for the MHD1 case, Fig. \ref{fig:diagonal} for MHD2,   Fig. \ref{fig:west} for MHD3 and Fig. \ref{fig:northfast} for MHD4. The coordinates of the points where the magnetic field is measured are indicated in each line for each of the cases.}
\end{figure}

 {\it Effects on the magnetic field.} We also study the effects of the initially uniform magnetic field, due to the shock cone formation process. The fact that the plasma piles up in a high density cone-shaped region, indicates that some important effects may happen, namely the magnitude of the magnetic field is expected to change and the resulting currents from the process of formation will promote Lorentz forces.
First we diagnose the magnetic field amplification. In Fig. \ref{fig:ampli1} we show the magnetic field strength as a function of time, measured at points where the plasma $\beta=\frac{2p}{b_i b^i}<1$, because there the magnetic field dominates over the hydrodynamical pressure, for the cases MHD1, MHD2, MHD3 and MHD4. In all these scenarios the highest amplification occurs at points near the black hole horizon. Notice that in the MHD2, MHD3 and MHD4 cases the magnetic field increases approximately between one and two orders of magnitude and in the case MHD1 the amplification is of one order of magnitude.

{\it Rarefaction regions and Lorentz force.} A second interesting implication of the distortion of the magnetic field is the formation of rarified zones not seen in the purely hydrodynamical case. Coincidentally, as shown for the cases MHD2, MHD3 and MHD4  in Figures  \ref{fig:diagonal}, \ref{fig:west}, \ref{fig:northfast}, these low density zones develop precisely where the magnetic pressure dominates over the fluid pressure $\beta<1$. In order to investigate the possible reason for this we tracked the Lorentz force $\epsilon_{ijk}J^j B^k$ where $J^j$ is the current density. In these Figures  we show a snapshot at  time $t=1000M$ (after the process has become stationary) of the rest mass density for the HD and MHD cases for comparison, and show the direction in which this Lorentz force acts. We observe that in the low density spots  the direction of the Lorentz force points in the appropriate direction as to move the plasma outwards. This is an indication that -even if an approximation- this force is the responsible for the formation of these spots. These rarefaction zones appear in different places depending on the direction of the wind, for example in the diagonal MHD2 case the rarified  zone is bigger in the top half of the shock cone but in the horizontal MHD3 case the zone is symmetric with respect to the $z=0$ plane. 
  
\begin{figure}
\centering
\includegraphics[width=4.15cm]{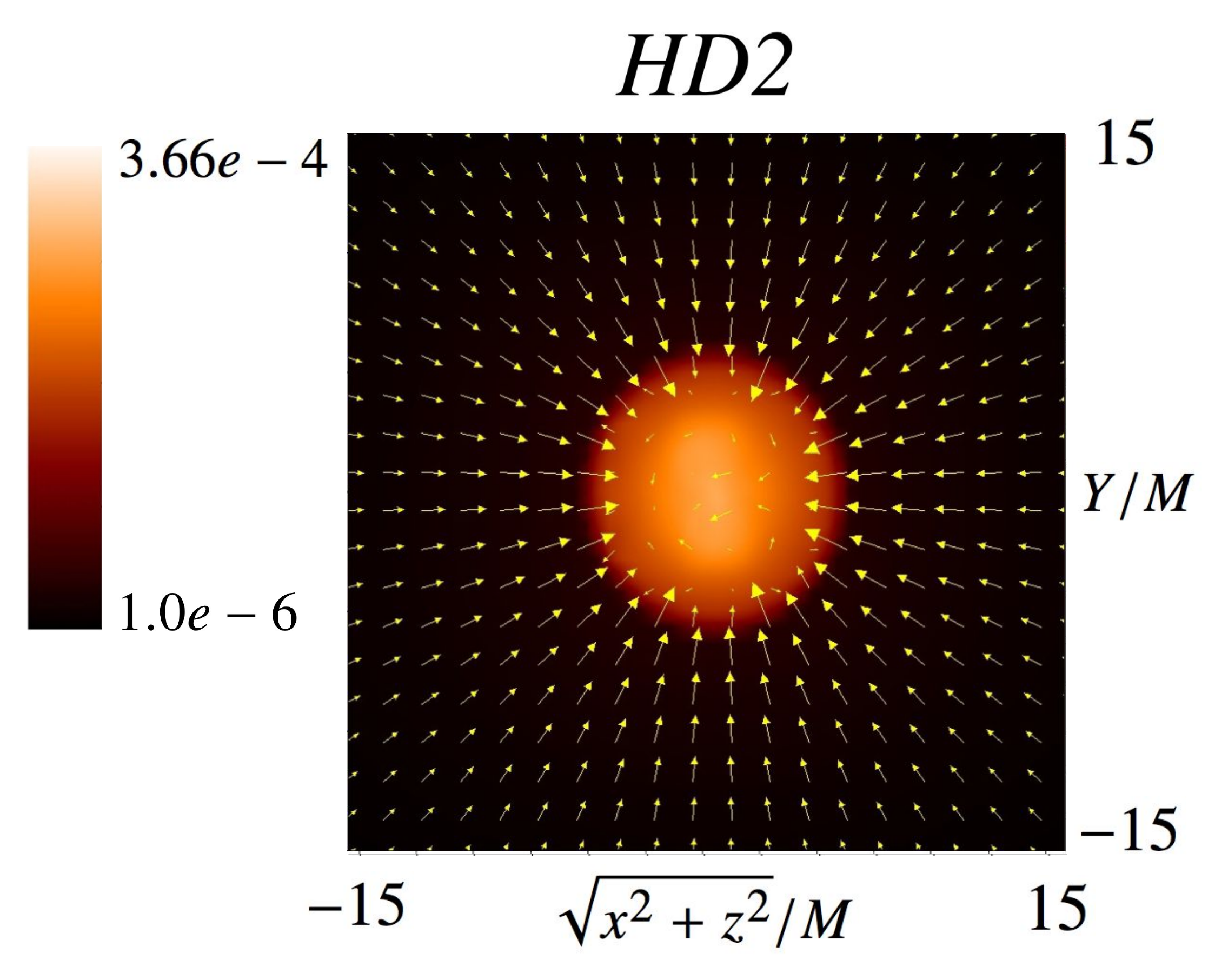}
\includegraphics[width=4.15cm]{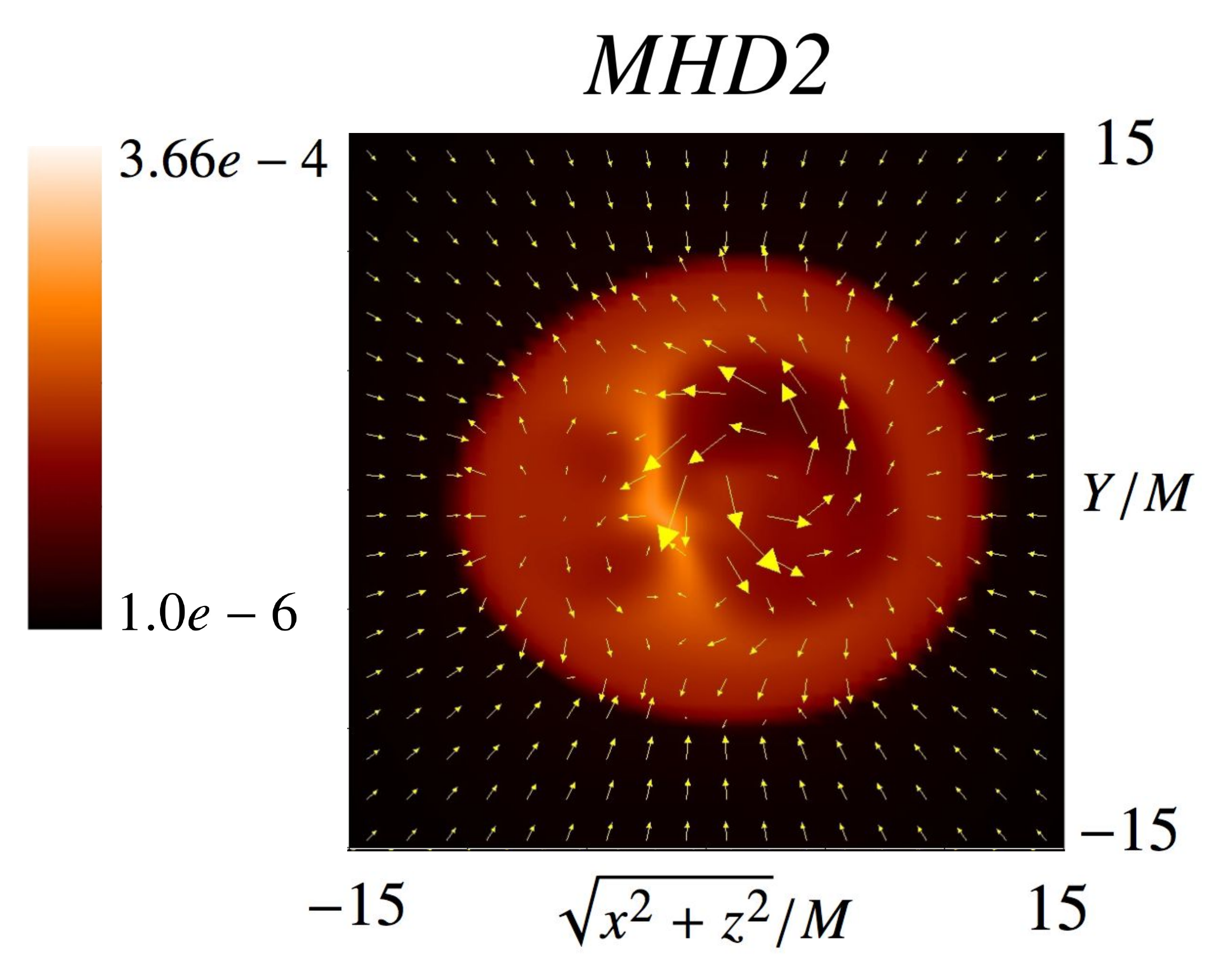}\\
\includegraphics[width=4.15cm]{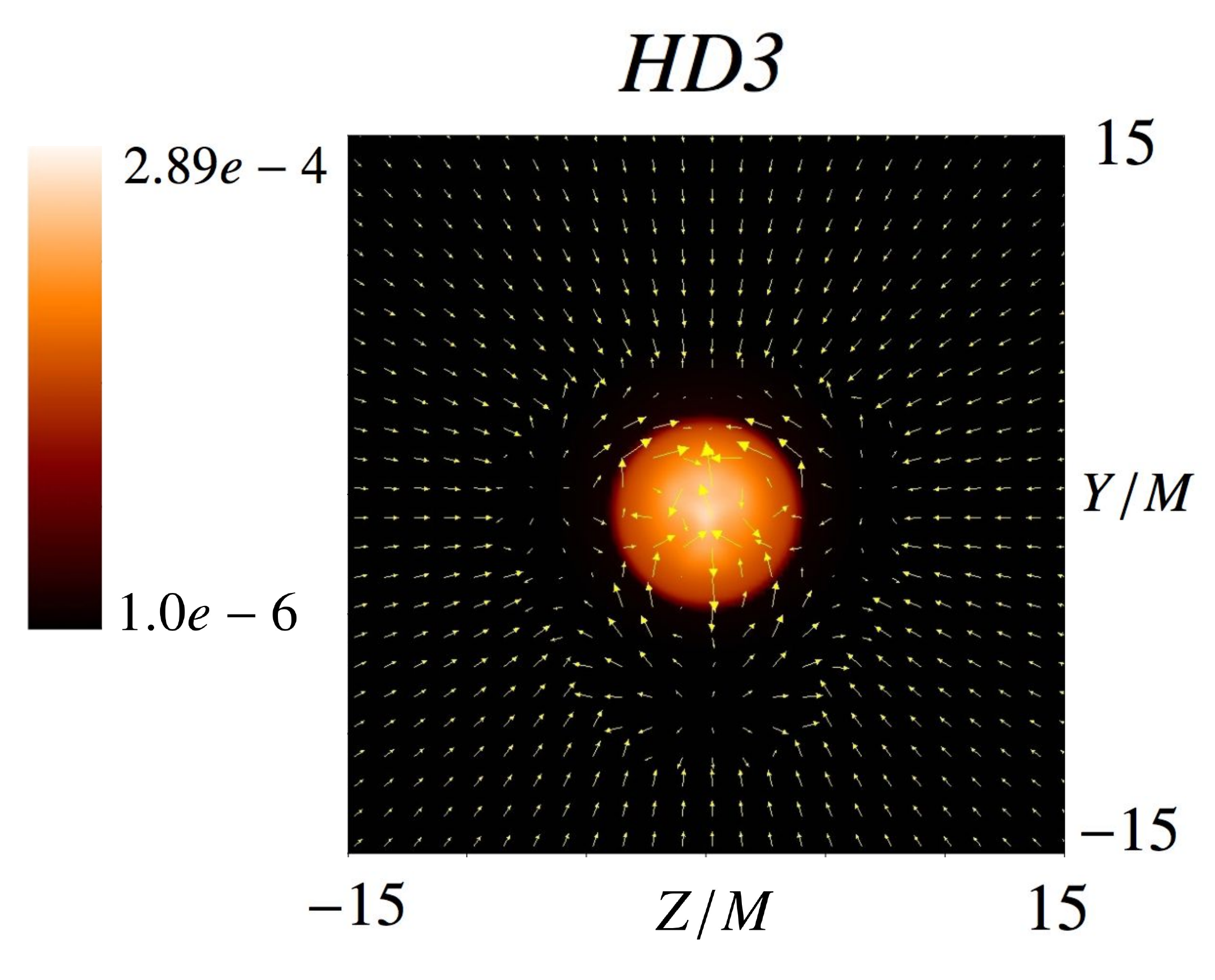}
\includegraphics[width=4.15cm]{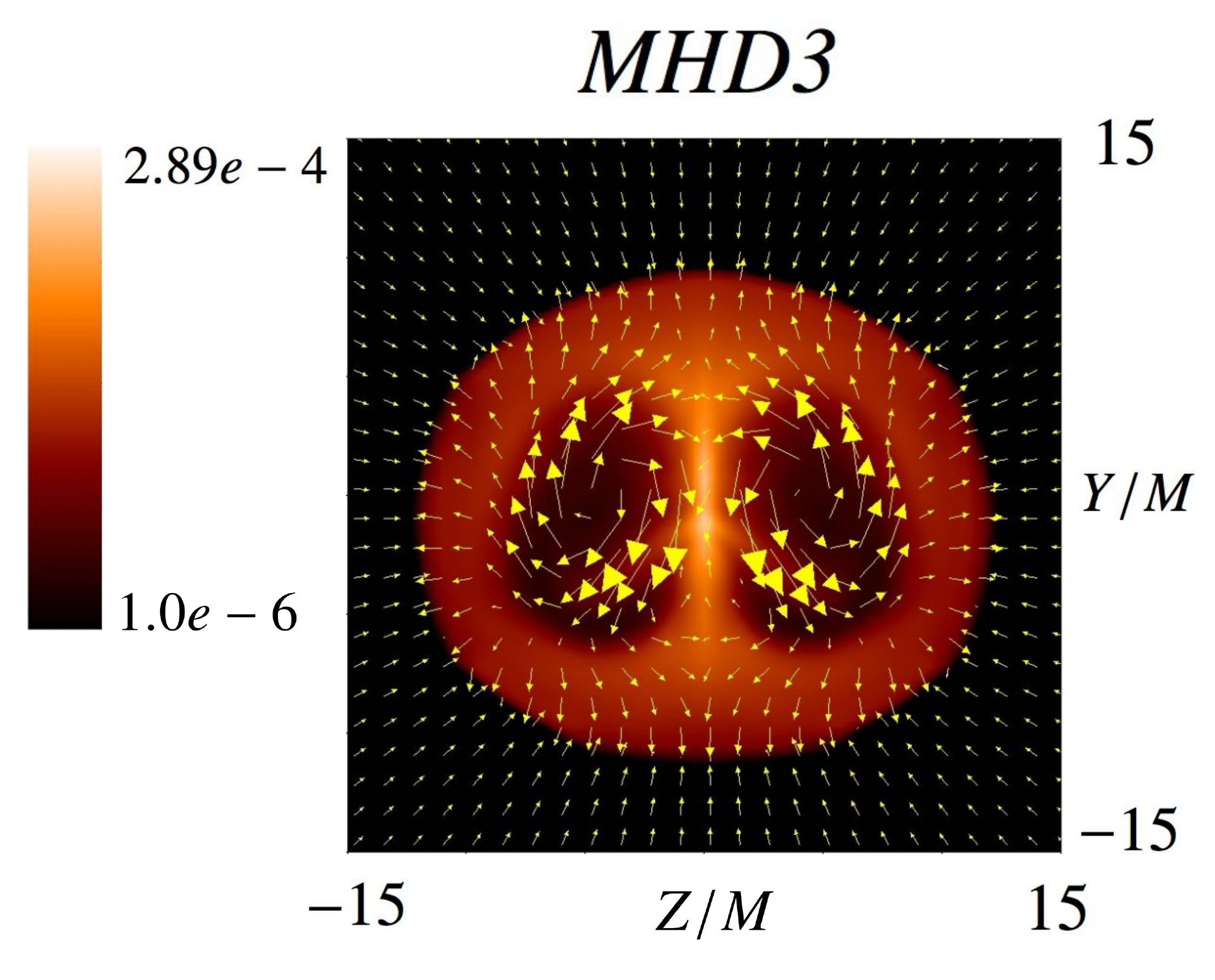}
\caption{\label{fig:vortex} 
Snapshot of the rest mass density $\rho$ and velocity field $v^i$ for models HD2, MHD2, HD3 and MHD3. For models HD2 and MHD2 we show the results on the plane $x+z=4$, which is a plane perpendicular to the wind direction. For models HD3 and MHD3 we show the results on the plane $x=-2$ which is a plane also perpendicular to the wind. The velocity field in both cases shows the formation of eddies precisely in the rarified zones of the shock-cone. It is useful to compare the shape and location of the rarified zones with the perpendicular view in Figures 2 and 3.}
\end{figure}

{\it Formation of eddies}. Another interesting effect is the formation of eddies in the shock cone.
In Fig. \ref{fig:vortex} we present the formation of eddies in two general cases, one is the diagonal case MHD2 $\nearrow \Uparrow$, and the second one is the horizontal wind MHD3 $ \Uparrow \leftarrow$.
For both cases we show the rest mass density on a plane perpendicular to the wind and near to the black hole, together with its purely hydrodynamical counterpart for comparison. We superpose the velocity field of the plasma in order to have a clear idea of the motion in the low density spots.

In the diagonal case HD2 and MHD2 we present the plot on the plane $x+z=4$, which is perpendicular to the wind in the shock-cone region at a close distance from the black hole's horizon. This map is a perpendicular view of that in Fig. \ref{fig:diagonal} and shows two different rarified spots that do not appear in the HD2 case. The most interesting part is that the velocity field indicates the plasma is rotating  approximately around the center of the big spot, whereas in the purely hydrodynamical case the gas is simply moving toward the center of the shock-cone.

In the horizontal wind case HD3 and MHD3 we show the density and velocity field of the shock-cone also in a plane perpendicular to the direction of the wind at a distance $2M$ from the center of the black hole. In this case, the symmetry allows to notice two symmetric low density spots. This is a perpendicular view from that in Fig. \ref{fig:west}. In this case it is clear that two eddies are formed in the rarified zones and the plasma is rotating around.

% ----- Table I
\begin{table}%[h!]
\caption{Parameters of the 8 simulations  used in this study. Four of them involve MHD and the other ones are the purely HD equivalent counterparts. In this Table the units of $a$ are $[1/M^2]$, the velocity  $v_{\infty}$ is in units of $c=1$ and the units of the magnetic field $B_0$ are $[1/M]$.}	
\begin{center}
\begin{tabular}{ccccc}\hline
Name&$a = 0.8$ &${v_{\infty}}=0.25$ & $B_0$ &Orientation\\\hline
MHD1&&& $1\times10^{-5}$ &$\uparrow \Uparrow$ \\\hline
HD1&&& $0$&$\uparrow \Uparrow$\\\hline
MHD2&& &$1\times10^{-5}$ & $\nearrow \Uparrow$ \\\hline
HD2&& & $0$ &$\nearrow \Uparrow $\\\hline
MHD3&&& $1\times10^{-5}$ &$ \Uparrow\leftarrow$ \\\hline
HD3&& & $0$&$ \Uparrow \leftarrow$\\\hline
MHD5&&& $1\times10^{-10}$ &$\uparrow \Uparrow$ \\\hline
MHD6&& &$1\times10^{-10}$ & $\nearrow \Uparrow$ \\\hline
MHD7&&& $1\times10^{-10}$ &$ \Uparrow\leftarrow$ \\\hline\hline
Name&$a = 0.8$ &${v_{\infty}}=0.5$ & $B_0$ &Orientation\\\hline
MHD4&& & $1\times10^{-5}$& $\uparrow \Uparrow$ \\ \hline
HD4&&& $0$ & $\uparrow \Uparrow$ \\ \hline
MHD8&& & $1\times10^{-10}$& $\uparrow \Uparrow$ \\ \hline
\end{tabular}
\label{tab:tabla1}
\end{center}
\end{table}

{\it Effect of the wind velocity.} One may wonder whether the wind velocity produces significant morphological changes. In the purely HD regime we found in our previous paper \citep{ours2015} that the accretion rate of the system and the open angle of the shock depends on the wind velocity, slower winds correspond to wider shock cones and higher accretion rates. In this paper we verified the same happens for the magnetized case. An important new feature appeared, this is the one related to the rarified zone within the shock-cone. The density in the rarified zone is one order of magnitude smaller in the fast case model MHD4 than in the slow case model MHD1 as can be seen in Figs. \ref{fig:northslow} and \ref{fig:northfast}.

{\it Accretion rate.} Another point to see is the influence of the magnetic field on the accretion rate. We measured $\dot M$ in a spherical surface located near the black hole's horizon, at $r=2M$ after the shock has been formed and the evolution regime becomes stationary. In Table \ref{tab:accretionrate} we show the values of the accretion rate for all the models at a stationary stage.  
We first observe that the horizontal wind (models HD3 and MHD3) has the highest accretion rate for the velocity used here, approximately between $10\%$ to $20\%$ above the other cases.

We found that the accretion rate of the MHD as compared with the purely hydrodynamical counterparts is within a $3\%$ of difference. This table also shows that the direction of the wind influences more the accretion rate than the fact of having pure HD or MHD. Besides the direction of the wind, another factor that influences the accretion rate is the wind velocity which can be seen from the comparison of models MHD1 and MHD4, with significant differences of 100\%.
 
% ----- Table II
\begin{table}%[h!]
\caption{\label{tab:accretionrate} Accretion rate values ($\dot M$) for all the cases. The perpendicular case to the axis of rotation of the black hole is the configuration with the highest $\dot M$. Moreover $\dot M$ does not change significantly between the MHD and HD cases.}	
\begin{center}
\begin{tabular}{ccccc}\hline
Model& Orientation &  $\dot { M }$ \\\hline
MHD1&$\uparrow \Uparrow  $ & $0.667 \times 10^{-3}$ \\\hline
HD1&$\uparrow \Uparrow$ & $0.682 \times 10^{-3}$\\\hline
MHD2&$\nearrow \Uparrow$& $0.673 \times 10^{-3}$\\\hline
HD2&$\nearrow \Uparrow$& $0.637 \times 10^{-3}$\\\hline
MHD3&$\rightarrow \Uparrow $ & $0.781 \times 10^{-3}$\\\hline
HD3&$\rightarrow \Uparrow $&$0.774 \times 10^{-3}$\\\hline\hline
Model& Orientation & $\dot { M }$ \\\hline
MHD4& $\uparrow \Uparrow$ &$0.315 \times 10^{-3}$\\ \hline
HD4& $\uparrow \Uparrow$ &$0.333 \times 10^{-3}$ \\ \hline
MHD5& $\uparrow \Uparrow$ &$0.679 \times 10^{-3}$\\ \hline
MHD6& $\nearrow \Uparrow$ &$0.635 \times 10^{-3}$ \\ \hline
MHD7& $\rightarrow \Uparrow$ &$0.769 \times 10^{-3}$\\ \hline
MHD8& $\uparrow \Uparrow$ &$0.331 \times 10^{-3}$ \\ \hline
\end{tabular}
\end{center}
\end{table}

\begin{figure}
\centering
\includegraphics[width=4.15cm]{Fig1a.pdf}
\includegraphics[width=4.15cm]{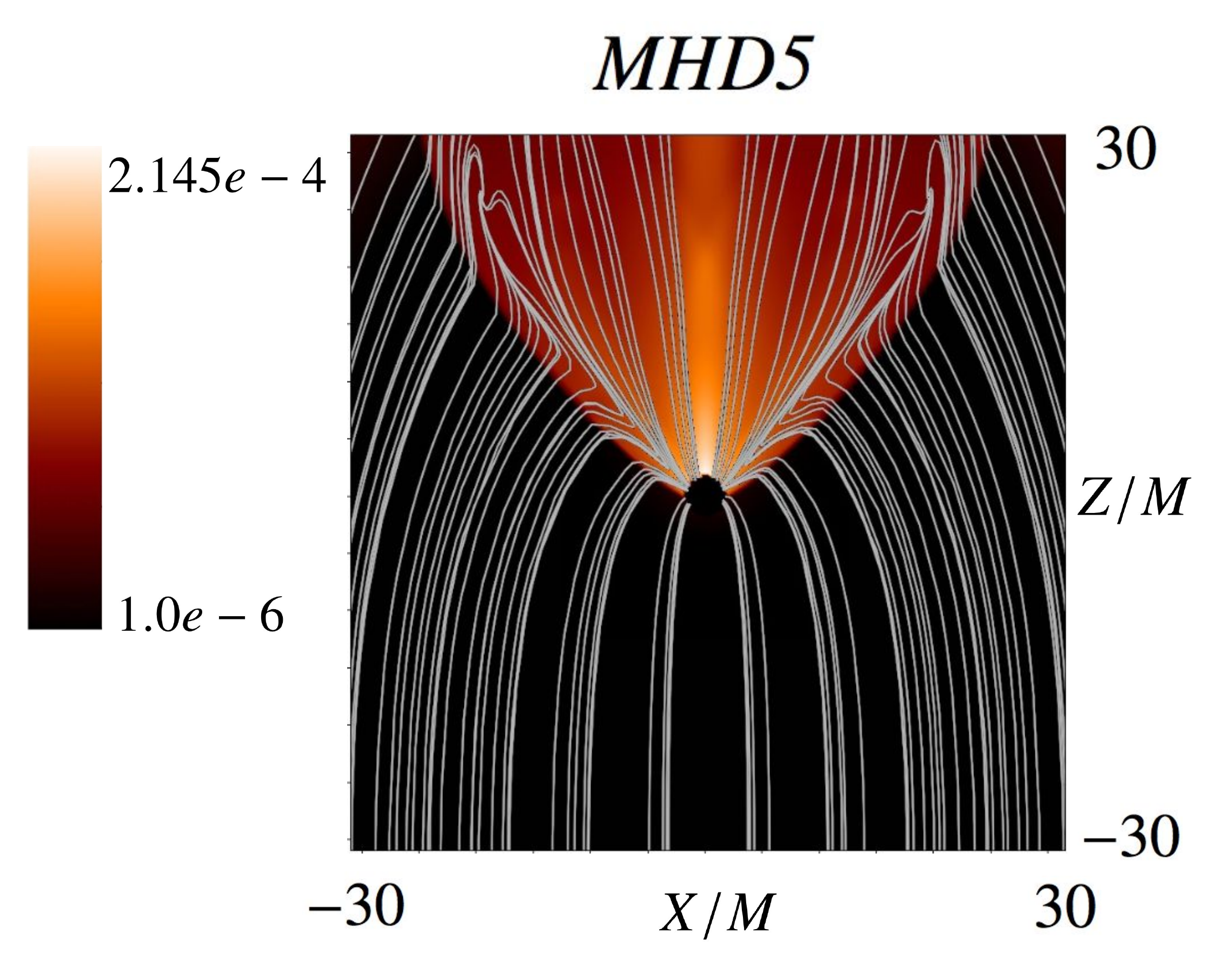}\\
\includegraphics[width=4.15cm]{Fig2a.pdf}
\includegraphics[width=4.15cm]{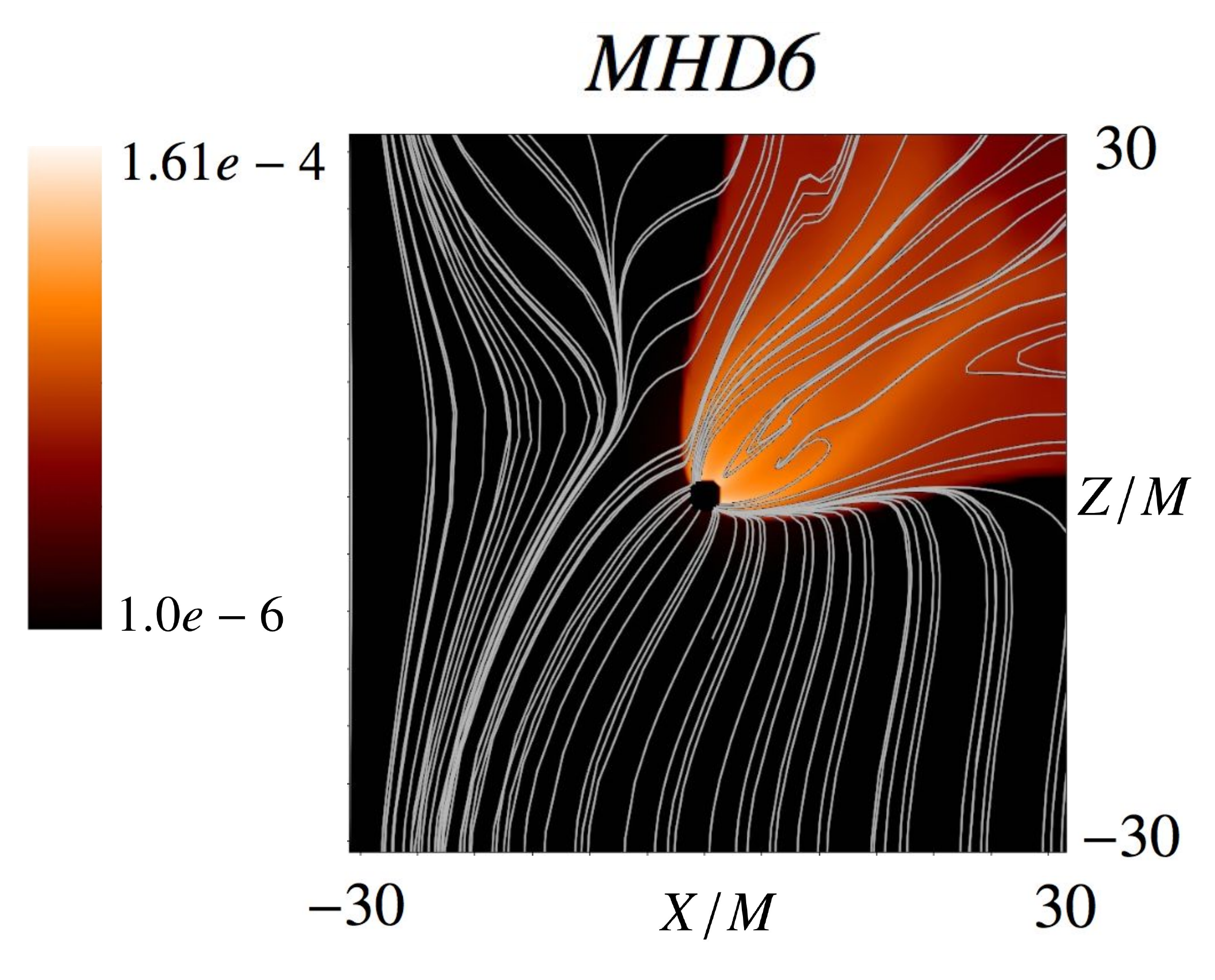}
\caption{\label{fig:bla1} 
Snapshot of the rest mass density $\rho$ during the stationary regime, for models HD1, HD2, and the MHD counterparts with the weak magnetic field strength $B_0=B_{0,weak}$, specifically MHD5 and MHD6. These models exhibit the distortion of the magnetic field lines likewise in the strong field case. However we did not find any drastic change in the morphology and magnitude of the rest mass density  that could for example trigger the formation of rarified zones.}
\end{figure}

\begin{figure}
\centering
\includegraphics[width=4.15cm]{Fig3a.pdf}
\includegraphics[width=4.15cm]{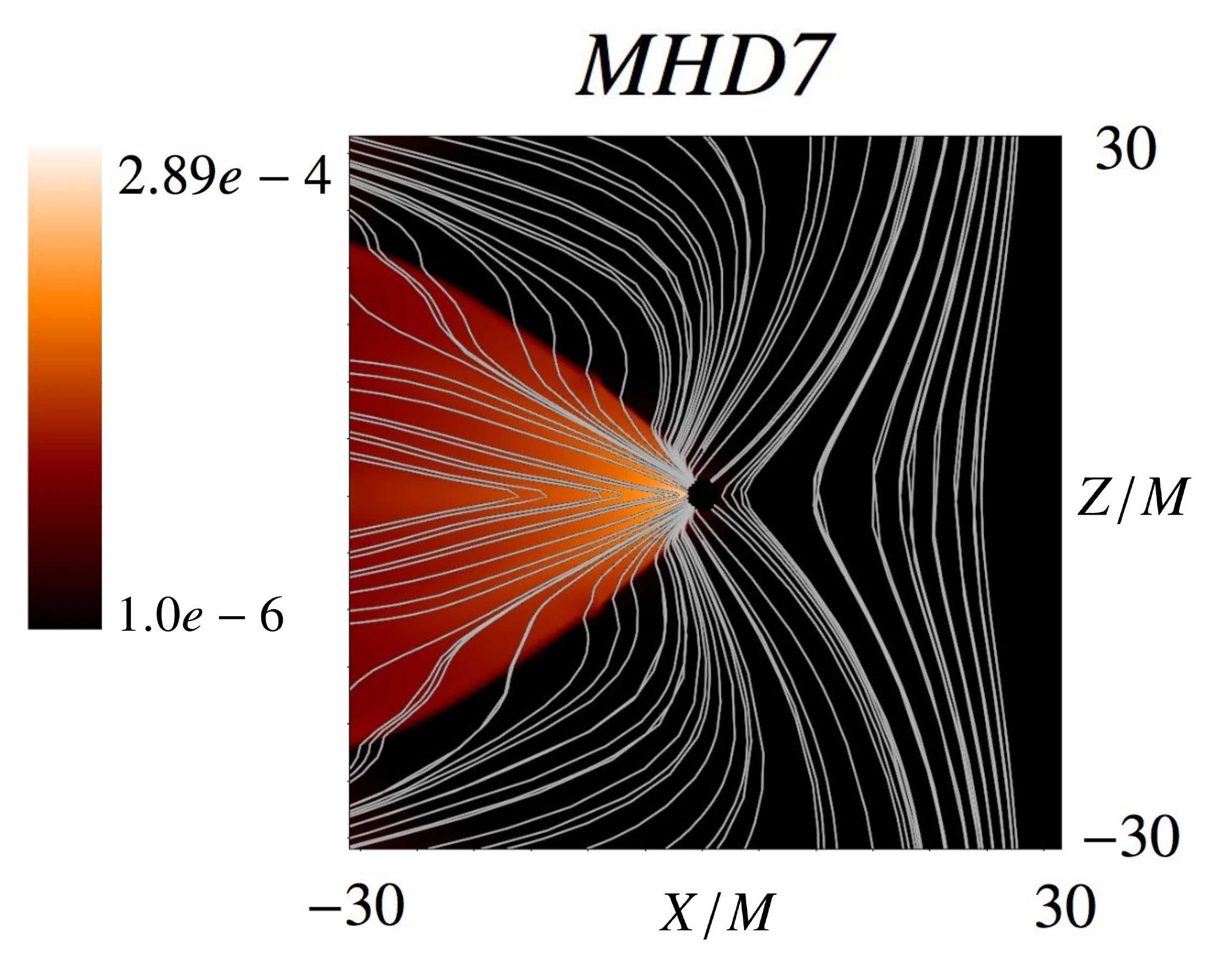}\\
\includegraphics[width=4.15cm]{Fig4a.pdf}
\includegraphics[width=4.15cm]{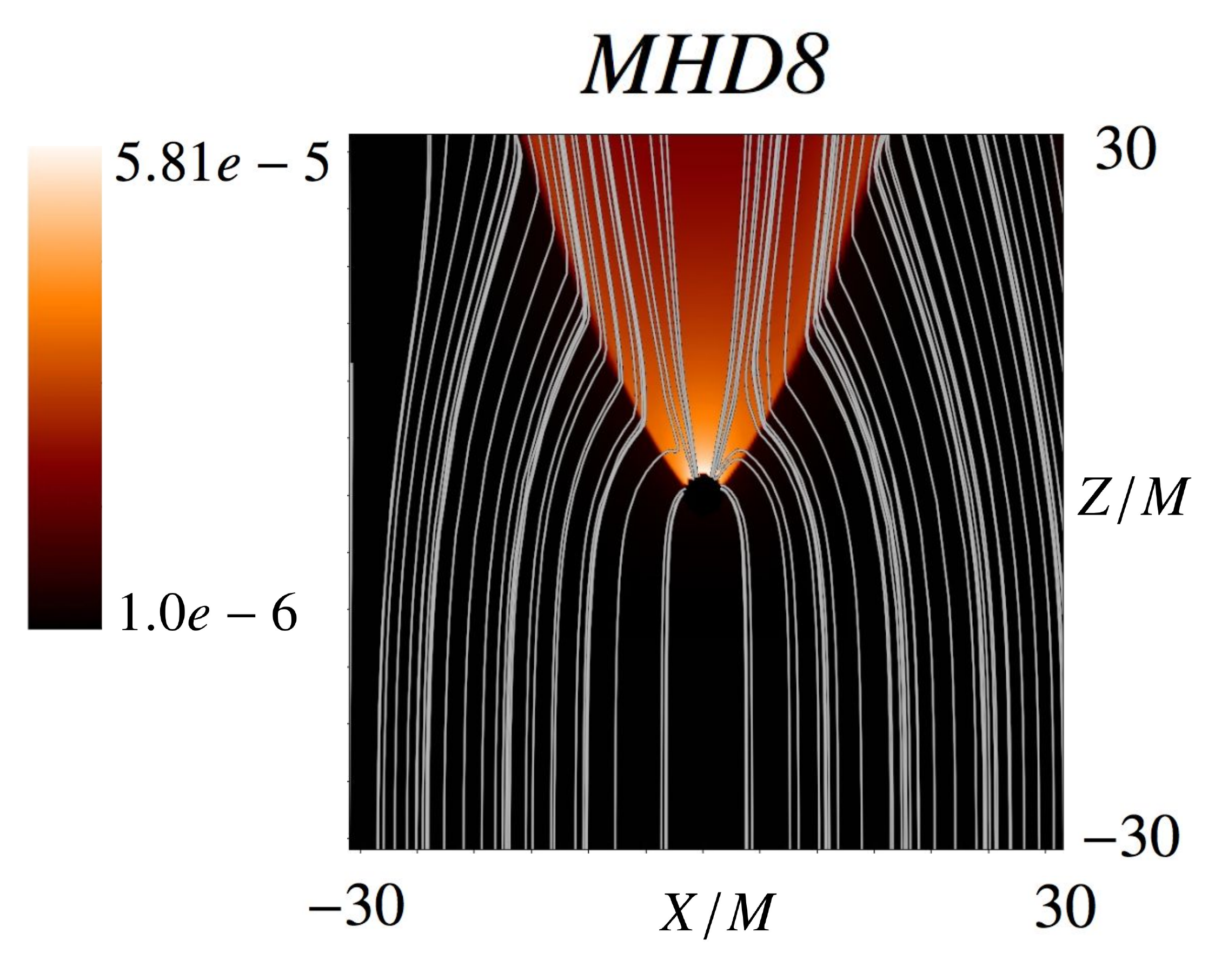}
\caption{\label{fig:bla2} 
Snapshot of the rest mass density $\rho$ during the stationary regime, for models HD3, HD4, and the MHD counterparts for the weak magnetic field $B_0=B_{0,weak}$, specifically MHD7 and MHD8. Again, we notice the distortion of the magnetic field lines but the field is not strong enough to produce zones of magnetic field domination that eventually could produce rarified zones.}
\end{figure}

{\it The weak magnetic field case $B_0=B_{0,weak}$.} Unlike the strong magnetic field case, the general properties for $B_0=B_{0,weak}$ (MHD5, MHD6, MHD7 and MHD8) are very similar to the purely hydrodynamical counterparts. In Figures \ref{fig:bla1} and \ref{fig:bla2}  we show the rest mass density of these models and the respective HD cases. Contrary to the strong magnetic field case, we do not observe a noticeable change in the shock cone morphology. We attribute this behavior to the fact that for this magnetic field the  magnetic field pressure is of order $10^{-20}$ and the plasma $\beta \gg 1$, even after the magnetic field is amplified. Thus there is no significant impact on the gas dynamics as the one observed in the strong field cases MHD1, MHD2, MHD3 and MHD4. In this sense this regime works as a correspondence case between MHD and HD, although the structure of magnetic field lines holds.

\begin{figure}
\centering
\includegraphics[width=8.5cm]{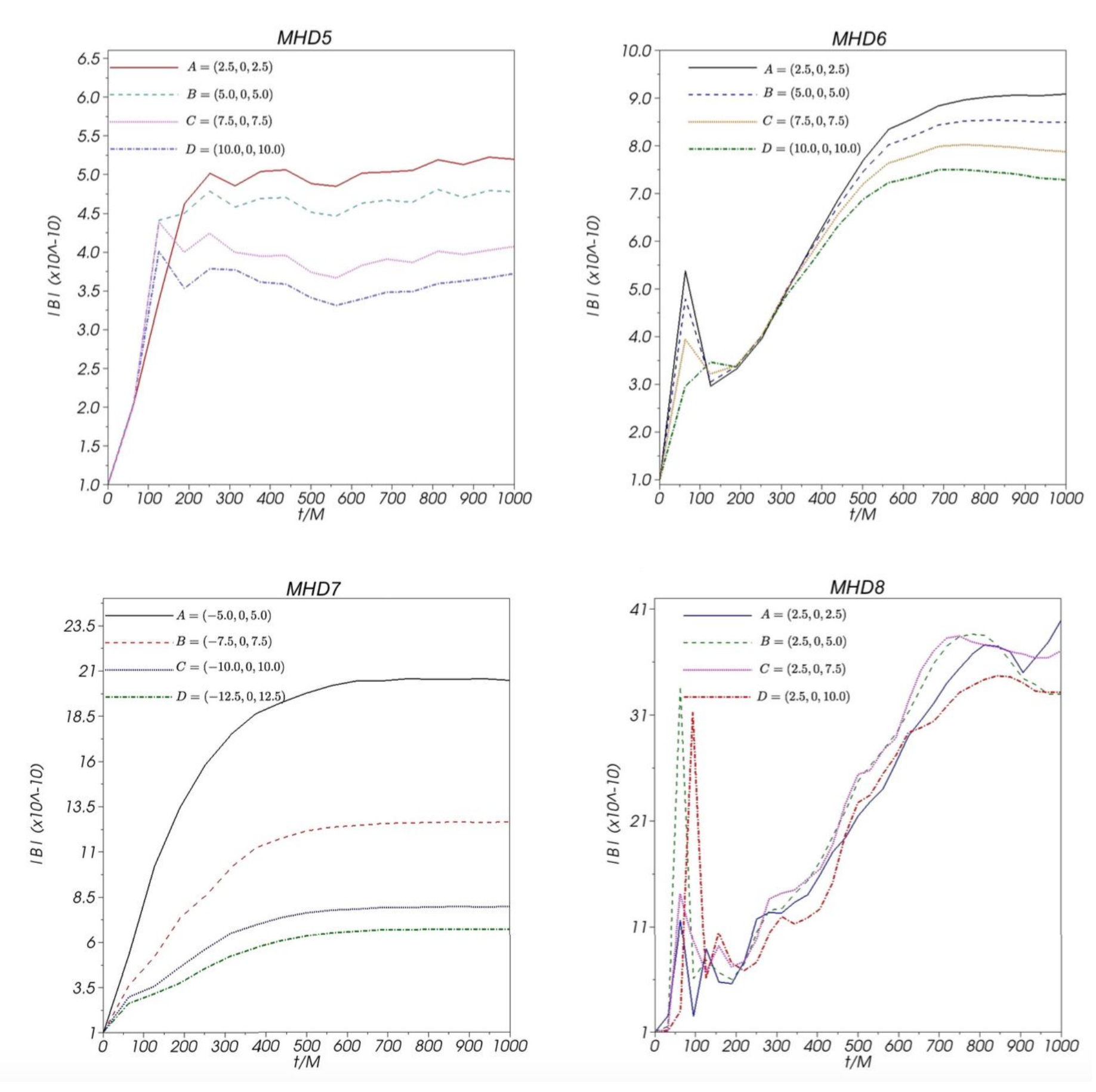}
\caption{\label{fig:ampli2} Time series of the magnetic field strength measured at four different points of the shock cone for $B_0=B_{0,weak}$ and models MHD5, MHD6, MHD7 and MHD8. For this we have chosen points located where the magnetic pressure dominates. These points can be identified in Fig. \ref{fig:bla1} for the MHD5 and MHD6 cases, and Fig. \ref{fig:bla2} for MHD7 and MHD8. The coordinates where the magnetic field is measured are indicated for each case.} 
\end{figure}

{\sl Effects on the magnetic field.} In Figure \ref{fig:ampli2} we show the amplification of the magnetic field, which is increased by  one order of magnitude. The initial and amplified values of the magnetic field are not sufficient to compete with the gas pressure, and this is the reason why there are not low density spots, first because the plasma $\beta$ is of the order of $10^{10}$ and second because the Lorentz force is ten orders of magnitude smaller than in the $B_{0,strong}$ case. 
As shown in Figures \ref{fig:bla1} and \ref{fig:bla2} this magnetic field is too small to produce any relevant change. In general, even though the magnetic field lines distort, purely hydrodynamics gas evolution rules the process.

{\sl Accretion rate.} Based on the previous description, for this scenario with a weak magnetic field, it is expected  the accretion rate to be even more similar to that of the HD counterparts. We find that the differences in accretion rate are within 0.5\%.

{\sl Attractor behavior.} The stationarity of the flow, morphology and magnetic field lines distribution is recovered when the wind density is varied with sinusoidal fluctuations of amplitude 10\% of $\rho_{\rm ini}$ and injected over a nearly arbitrary time window as long as $\rho_{\rm ini}$ is recovered. This indicates that the configurations resist inhomogeneities, which are expected to happen in real scenarios. More formally, it would be interesting to find a parametrization of the basin of attraction of these nearly-stationary accretion configurations, and how rapidly they approach stationarity in terms of Lyapunov exponents.

% ---------------------------------------------
% ---------->     SECTION     <----------
% ---------------------------------------------

\section{Conclusions and Discussion}
\label{sec:conclusions}

We present the  accretion of a supersonic magnetized wind onto a black hole in 3D. For this we selected a set of different wind orientations with respect to the spin of the black hole. In all the cases we compared the results with the purely hydrodynamical counterpart.

We studied two different values of the magnetic field strength, strong and weak cases. We found that in the strong field case there are zones within the shock cone in which the magnetic field pressure dominates $\beta<1$ and produces low density spots within the shock cone. We also show the formation of eddies  consisting of plasma rotating around the low density spots within the shock cone near the black hole horizon. Using an approximate calculation of the Lorentz force, we found that this force can be the responsible for the depletion of the low density regions.
The weak field case on the other hand, was such that the gas pressure dominates over the magnetic field and the plasma $\beta \gg 1$, and neither low density spots nor eddies are formed, although the magnetic field lines follow a similar pattern as in the strong field case.

We find that the accretion rate is not significantly different between the HD and MHD cases. Instead, the wind velocity and orientation is more important  for the accretion  rate.

We are sure that with the insight obtained with the general scenarios presented here, the infrastructure developed, specifically the implementation of upstream and downstream boundary conditions, can be  applied to astrophysical systems, like wandering black holes of type QSO 3C186 \citep{chiaberge2017} and supernovae natal kicks like that in  \citep{https://doi.org/10.48550/arxiv.2201.13296}. The various properties of the process, including the morphology of the shock cone, the formation of rarified zones and eddies are expected to change for different  parameters of the wind, including the asymptotic velocity and equation of state of the gas. It is possible now to construct a catalog of simulations with astrophysical parameters to be contrasted with observations. We also expect that our set up, applied to specific scenarios will be useful for example in the study of the cases like the high speed black hole prior to merger, during the common envelop face in \citep{Cruz_Osorio_2020}.
  
Concerning the observability of the processes detailed in this paper, a direct observation does not seem to be affordable at the moment, considering the candidates need a higher resolution than those used by the EHT. Then it is not expected a direct observation of the shock-cone itself, and differences with the presence of magnetic fields should be even finer. However it can be expected that the formation process of the shock-cone will reveal the intrinsic properties of the system, like black hole velocity and mass, perhaps spin, and the gas density and equation of state as illustrated in \citep{GonzalezGuzman2018} for the case of a wandering black hole, along with a correlation of vibrational modes of the shock-cone as indicated in \citep{Lora_Clavijo_2013}. Vibrations, that will depend on each combination of parameters of wind inclination angle and magnetic field strength, that we showed here that have local fingerprints, should be analyzed as done for the spin-less, hydrodynamical model in \citep{Lora_Clavijo_2013}. The various scenarios, with the appropriate vibration modes would compose the catalog of possible scenarios to be correlated with observations on candidates like QSO 3C186. On aother case, the binary black hole merger, the first steps have been already explored and signatures are expected together with gravitational wave signals \citep{Cruz_Osorio_2020}.

% ----->     ACKNOWLEDGMENTS     <-----

\section*{Acknowledgements}

MGL is supported by NSF grant PHY-1550461, PHY-2207780, PHY-2114582, and the Mexican National Council of Science and Technology (CONACyT) CVU 391996. FSG acknowledges support from grant CIC-UMSNH-4.9. The runs were carried out in the Big Mamma cluster of the Laboratorio de Inteligencia Artificial y Superc\'omputo, IFM-UMSNH.

% ----->     ACKNOWLEDGMENTS     <-----

\section*{Data availability}

The data underlying this article will be shared on reasonable request to the corresponding author.\\

% ---------------------------------------------
\bibliographystyle{mnras}

\bibliography{MHDWinds}

% ---------------------------------------------
% ---------->     SECTION     <----------
% ---------------------------------------------
\appendix
\section{Convergence}

%\subsection{Convergence}

The implementation of the relativistic hydrodynamics  GRHydro thorn, both in special and general relativity of the ETK has been tested severely for particular test scenarios \citep{ETK2012,baiotti2005}. In order to support the validity of our results we present a self-convergence test for one of the scenarios described in this paper. We show the order of convergence  $Q$ for the density along the $z$ axis outside of the black hole horizon, where the high density shock-cone forms,  at two different times $1000M$ and $2000M$ for the MHD4 case. 

\begin{figure}
\includegraphics[width=8.5cm]{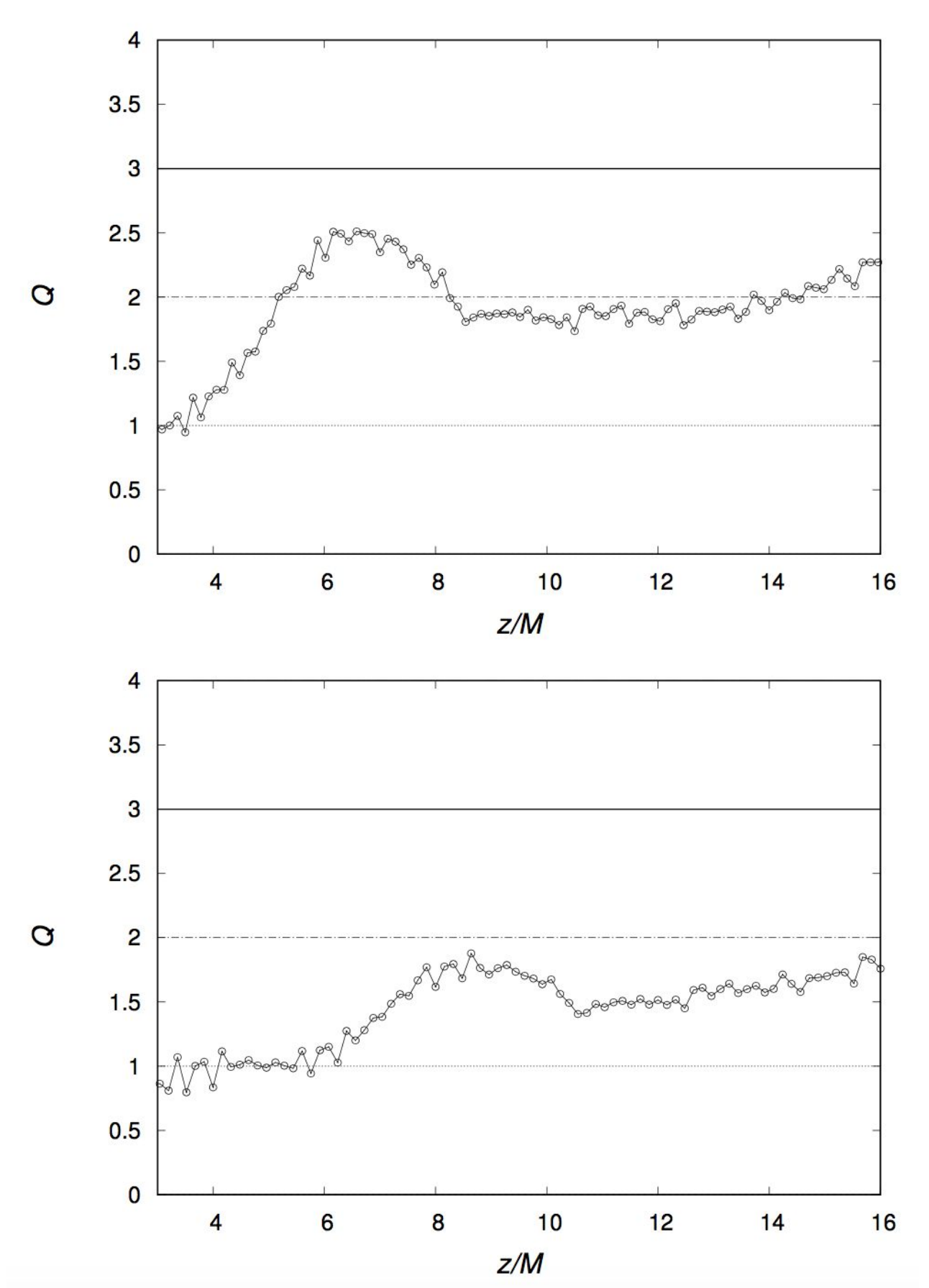}	
\caption{\label{fig:convergence} Order of convergence $Q$ of the rest mass density measured along the $z$ axis, on the downstream zone, at two different times, $t=1000M$ (top) and  $t=2000M$ (bottom) for the case MHD4 along the shock cone.}
\end{figure}

For this test we calculated three numerical solutions $\rho_1$, $\rho_2$ and $\rho_3$ using respectively the resolutions $\Delta x_1=0.5M$, $\Delta x_2=\Delta x_1/2$ and $\Delta x_3=\Delta x_2/2$.  We then evaluate $Q=\frac{\log{(\rho_1 -\rho_2)}-\log{(\rho_2 -\rho_3)}}{\log{2}}$ and show its value in Figure  \ref{fig:convergence}. The result is that $Q$ takes values between 1 and 2, which is the expected order of convergence of the high resolution shock capturing methods using the HLLE flux formula, with a linear reconstructor in the presence of shocks. This shows that the simulations for the evolution of the wind, and for the parameters in this paper, are carried out within the convergence regime of the numerical methods used.

\end{document}